\documentclass[preprint, nonatbib]{elsarticle}

\usepackage{graphicx}
\usepackage{caption}
\usepackage{soul}
\usepackage{color}
\usepackage{placeins}
\usepackage{xcolor}
\usepackage{subcaption}
\usepackage{amsmath}
\usepackage{textcomp, gensymb}
\usepackage{hyperref}
\usepackage{lineno}

\DeclareUnicodeCharacter{03BC}{$\mu$}

\makeatletter
\let\c@author\relax
\makeatother

\usepackage{biblatex}

\bibliography{references}

\journal{Flow Measurement and Instrumentation}

\begin{document}

\title{In-Lab X-ray Particle Velocimetry for Multiphase Flows: Design Principles and Demonstration of \textit{O}(1~kHz) XPV}

\author[1]{Jason T. Parker}
\ead{jtparker@berkeley.edu}

\author[2]{Till Dreier}

\author[2]{Daniel Nilsson}

\author[1]{Simo A. M\"akiharju}


\affiliation[1]{organization = {Department of Mechanical Engineering, UC Berkeley},
addressline = {2521 Hearst Ave.},
city = {Berkeley},
citysep={},
postcode = {94720},
state = {CA},
country = {USA} }

\affiliation[2]{organization = {Excillum AB},
addressline = {Jan Stenbecks Torg 17},
city = {Kista},
citysep={},
postcode = {164 40},
country = {Sweden} }

\begin{abstract}
We combine X-ray-specific tracer particles, a photon counting detector, and a liquid metal jet anode X-ray source to achieve \textit{O}(1~kHz) X-ray imaging speeds in the laboratory, 15$\times$ faster than previous comparable studies with \textit{O}(50~\textmu m) tracers. To examine the limits of this measurement technique we conduct three experiments: 2D and 3D X-ray particle velocimetry (XPV) of Poiseuille pipe flow, 3D XPV of flow around a Taylor bubble, and 3D scalar mixing with a laminar jet. These experiments demonstrate the performance improvement achievable by combining the aforementioned elements, the applicability to multiphase flows and deforming systems, and the potential to capture scalar and vector quantities simultaneously. Most importantly, these experiments are conducted with a laboratory-scale system, showing that in-lab X-ray particle velocimetry techniques are now becoming usable for a wider range of flows of interest. Furthermore, the design of XPV experiments is discussed to clarify the trade offs between achievable imaging speed, domain size and spatiotemporal resolution. 
\end{abstract}

\maketitle

\section{Introduction} \label{sec:intro}
Particle tracking velocimetry and particle image velocimetry are commonly used in contemporary fluid dynamics experiments. One of the major limitations of these techniques, though, is their inability to measure flows that are opaque to visible wavelengths of light. The opacity could be due to either the surrounding media, the fluid itself, or multiple refractive interfaces. X-rays, on the other hand, have an index of refraction near unity, and many materials are transparent to X-rays. Recently, researchers have started to combine X-ray imaging and particle-based velocimetry to study multiphase flows \cite{aliseda_x-ray_2021, makiharju_dynamics_2017, makiharju_time-resolved_2013, ganesh_bubbly_2016, heindel_x-ray_2008, seeger_x-ray-based_2001, rodriguez_effect_2023}, porous media flows \cite{makiharju_tomographic_2022, bultreys_x-ray_2022, kingston_characterizing_2015, gollin_performance_2017}, biological flows \cite{antoine_flow_2013, jamison_x-ray_2012, kim_x-ray_2006, park_x-ray_2016}, and many others that are difficult to examine with visible light image-based measurement techniques. So far, however, the temporal resolution of in-lab X-ray velocimetry and other X-ray imaging-based techniques has limited the range of flows that can be studied. Wider availability of enhanced X-ray particle tracking velocimetry (XPV) and particle image velocimetry (XPIV) could offer previously unobtainable insight into many flows of interest.

Unlike their visible light counterparts, 2D and tomographic XPTV and XPIV -- all of which comprise a class of measurement that can be more generally called X-ray Particle Velocimetry (XPV) -- have not been capable of reaching frame rates \textit{O}(1-100~kHz) in the laboratory with \textit{O}(50~\textmu m) tracers needed for many applications. Such imaging speeds may be possible at synchrotrons, but with limited beam time availability and typically only \textit{O}(1~mm) fields of view, synchrotrons severely restrict flow experiments.

There are a couple primary reasons for limited in-lab XPV. Although continuously improving, X-ray imaging detectors are generally orders of magnitude slower than the most advanced visible light high-speed cameras. Even if X-ray imaging detectors were fast enough, the relatively modest polychromatic photon flux generated by most in-lab X-ray sources imposes another limitation. A brighter X-ray source is necessary to enable short exposure times while retaining usable image quality. As laboratory X-ray sources are orders of magnitude dimmer than a synchrotron, much of the prior work in XPV has been done at synchrotrons, which can achieve photon fluxes order of \textit{O}($10^{17}$)~ph/m$^2$/sec \cite{macdowell_x-ray_2012}. 

Lee and Kim \cite{lee_x-ray_2003} captured the first XPV measurement at the Pohang Light Source twenty years ago. They generated the image contrast by attenuation, with edge enhancement due to refraction, and measured the 2D-projection of Poiseuille flow. Subsequently, Im et al. \cite{im_particle_2007} employed phase contrast imaging at the Advanced Photon Source, and used tomography to reconstruct the velocity field of pipe flow in 3D. Measurements of canonical low speed flows eventually evolved to more complex high speed flows such as the cavitating sheet experiment recently performed by Ge et al. \cite{ge_synchrotron_2022} at the Advanced Photon Source synchrotron. Alas, while incredibly bright, synchrotrons impose practical limitations on the types of experiments one can run. Typically, synchrotrons beam widths are on the order of a few millimeters, making large experiment geometries impossible. Multiphase flows in particular are notoriously challenging to scale because complete similarity is rarely achieved in scaled multiphase flow experiments. Furthermore, synchrotron beam time is in finite supply and high demand. For most users, using a synchrotron requires costly, inconvenient travel. When one does get time on synchrotrons, it is often for a few hours to a few days, making long-term or complicated experiments challenging to conduct.

The limitations posed by synchrotrons motivate research to develop XPV techniques that are practical for laboratory use. Prior studies demonstrated that XPV is feasible in a laboratory setting with currently available equipment \cite{parker_experimentally_2022, parker_enhanced_2022, makiharju_tomographic_2022, lappan_x-ray_2020}. However, most of these experiments use exposure times \textit{O}(10-100~ms) – too slow for most flows of interest. Many prior studies also use large, high Stokes number particles \cite{heindel_x-ray_2008, lappan_x-ray_2020, drake_developing_2012}. These were appropriate for the selected application, but large particles severely limit the kind of flows that could be examined. For energetic or low-speed flows they may not be accurate flow tracers due to inertial and buoyancy effects, respectively.

The three primary limitations for in-lab XPV currently are detector performance, tracer contrast, and source brightness \cite{makiharju_tomographic_2022}. The means to address these shortfalls work best in combination: ``noiseless" photon counting detectors (PCDs) \cite{russo_handbook_2018} with new, high contrast, neutrally buoyant flow tracers \cite{parker_enhanced_2022}, and brighter in-lab sources such as liquid metal jet (LMJ) anode sources \cite{hemberg_liquid-metal-jet_2003}. Previous work has demonstrated the effectiveness of the former two approaches \cite{parker_enhanced_2022}. This study combines for the first time a LMJ X-ray source with X-ray imaging-specific tracer particles and a PCD, and further discusses the overall design of in-lab XPV experiments. 

Laboratory-scale LMJ X-ray sources can be orders of magnitude brighter than typical laboratory solid target sources for a given focal spot size \textit{O}(10~\textmu m). A jet of liquid metal acts as the anode, overcoming the thermal limitations of stationary or rotating anode sources. That is, LMJ sources can achieve orders of magnitude higher output power density, achieving greater brightness without defocusing the focal spot, which would result in blurring and thus limit the spacial resolution.

For this study, we use an Excillum MetalJet E1+ X-ray source to capture up to 1~kHz X-ray footage of \textit{i}) laminar pipe flow, \textit{ii}) Taylor bubble flow, and \textit{iii}) laminar jet scalar mixing. We demonstrate that LMJ sources, coupled with a PCD and contrast-optimized nominally neutrally \textit{O}(50~\textmu m) tracer particles, enable particle tracing in \textit{O}(10~mm) domains with exposure times roughly 15$\times$ lower than previously achieved. 1~kHz frame rates enable domain-resolving full 360 degree computed tomography (CT) up to 3~Hz, limited in this case by the speed of the rotation stage.

These advancements make resolving flow with speeds \textit{O}(100~mm/s) possible with an \textit{O}(50~\textmu m) tracer particle. Such flow speeds are well within the range of human blood flow \cite{klarhofer_high-resolution_2001} or flows due to natural convection relevant to vitrification \cite{rabin_mathematical_2021, parker_direct_2023}, for example. Many porous media flows fall into this category as well \cite{makiharju_tomographic_2022, bultreys_x-ray_2022}. In effect, this study demonstrates that laboratory XPV is currently usable for studying previously inaccessible flows of interest.

This paper is organized as follows: section \ref{sec:expdes} details the flow experiment setup, X-ray source, tracer particles, and detector; section \ref{sec:resdic} discusses the experiment results; section \ref{sec:design} discussed in-lab XPV experiment design; finally, section \ref{sec:conc} provides the conclusions and future outlook for laboratory XPV. Additional details on image processing and particle tracking algorithms are provided in the appendices.

\section{Materials and Methods} \label{sec:expdes}
\subsection{Flow Setup}
Three experiments are conducted with one setup to explore what is achievable with a brighter X-ray source in combination with improved flow tracers and a PCD. As discussed in section \ref{sec:intro}, many of the early synchrotron XPV experiments that focused on technique development measured developed pipe flow. It is straightforward to compare to theory; we replicate these experiment as the first of the in-lab XPV experiments for this study. This also enables clear comparison to \cite{parker_enhanced_2022}, which used an identical pipe, fluid, and nearly identical tracer particles.

We first study developed pipe flow with both 2D-projected and tomographic XPV (TXPV). The second experiment considers flow around a Taylor bubble, the results from which can be compared to \cite{makiharju_tomographic_2022}, which used a similar setup but a traditional solid-target X-ray source, scintillator-based detector and silver coated hollow tracer particles (commonly used for visible wavelength light PIV). Lastly, we explore the systems ability to quantify 3D mixing of a potassium iodide (KI)-water solution jet injected into glycerine. One of the goals of in-lab XPV is to be able to conduct high-speed CT reconstructions of the domain whilst also measuring fluid velocity via XPV. The KI jet experiments are used to test the limits of the high-speed CT reconstructions with this system.

\begin{figure}
    \centering
    \includegraphics[width = 0.99\textwidth]{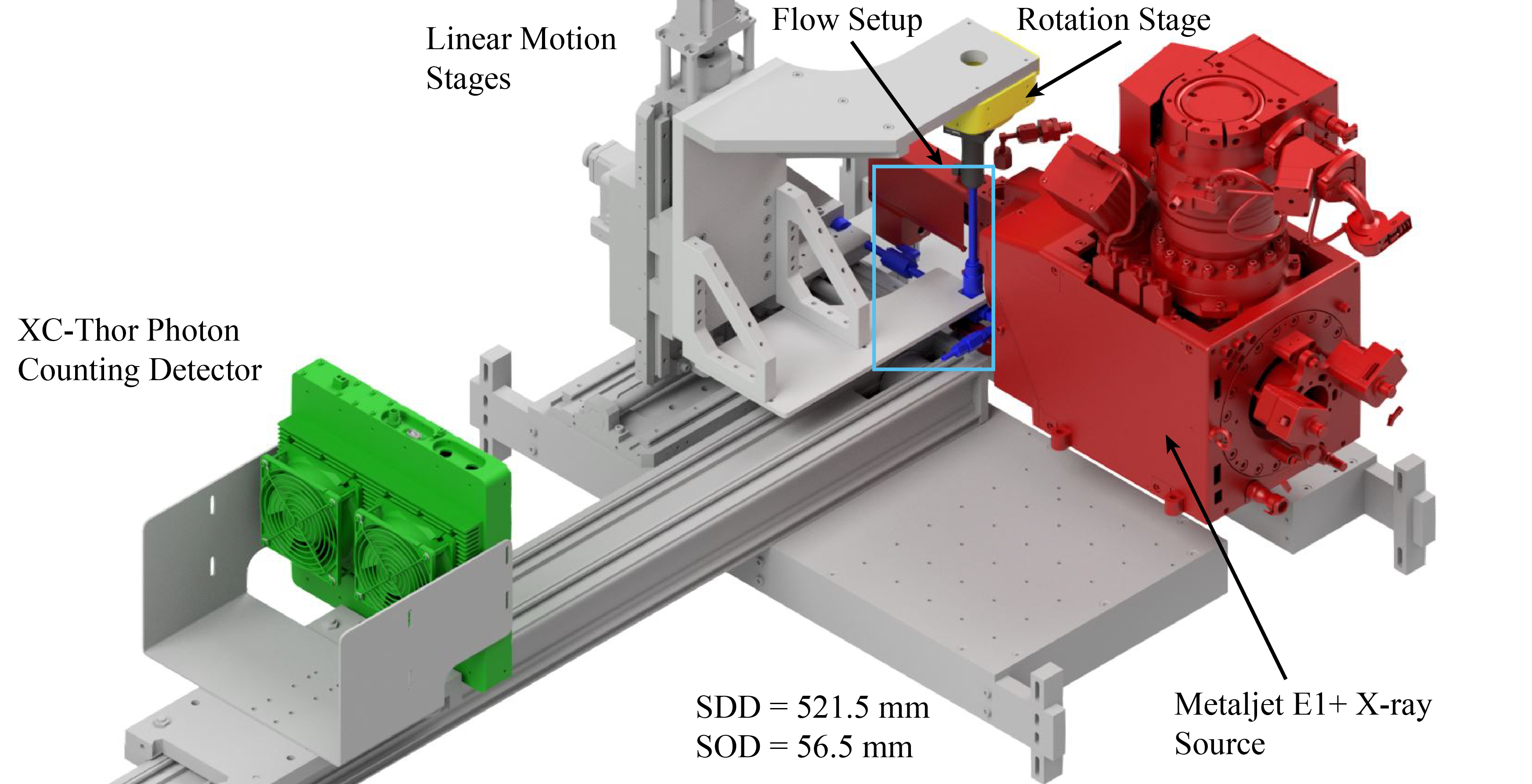}
    \caption{The X-ray imaging setup. The open flow loop is placed in front of the X-ray source aperture. A LAB Motion RT100S air bearing rotation stage controls rotation of the pipe which is connected to a syringe pump via a slip-joint allowing smooth continuous pipe rotation while the piping upstream remains fixed.}
    \label{fig:imgsetup}
\end{figure}

Figure \ref{fig:imgsetup} depicts the X-ray imaging setup. For these experiments, the source-to-object distance (SOD) is 56.5~mm; the source-to-detector distance (SDD) is 521.5~mm. We measure the former from the focal spot to the center of the pipe; the latter from the focal spot to the face of the detector pixel plane. The field of view in object-plane is 22.2~mm $\times$ 2.8~mm due to geometric magnification.

\begin{figure}
     \centering
     \begin{subfigure}{0.65\textwidth}
         \centering
         \includegraphics[width=\textwidth]{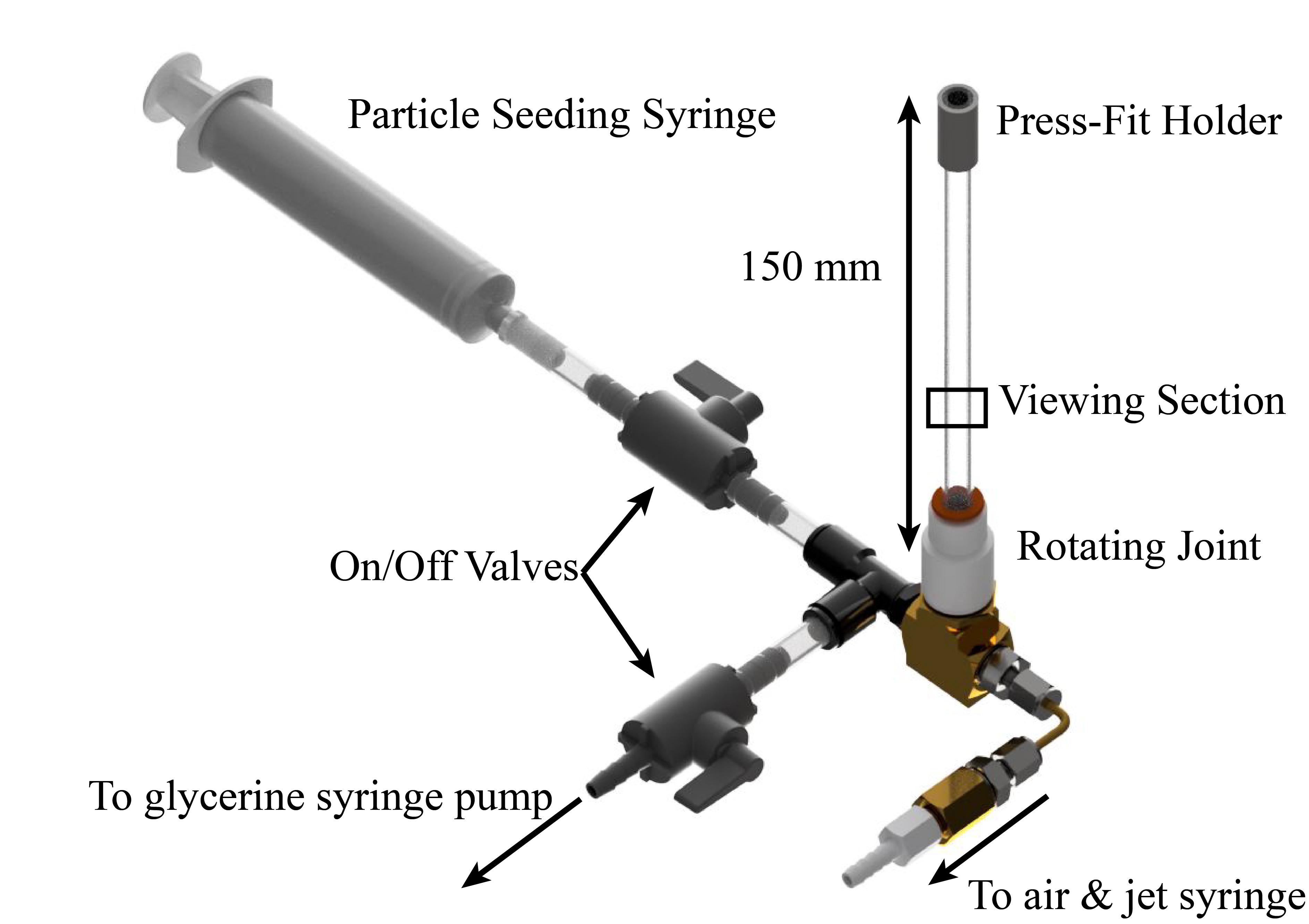}
         \caption{}
         \label{fig:flowsetup1}
     \end{subfigure}
     \hfill
     \begin{subfigure}{0.65\textwidth}
         \centering
         \includegraphics[width=\textwidth]{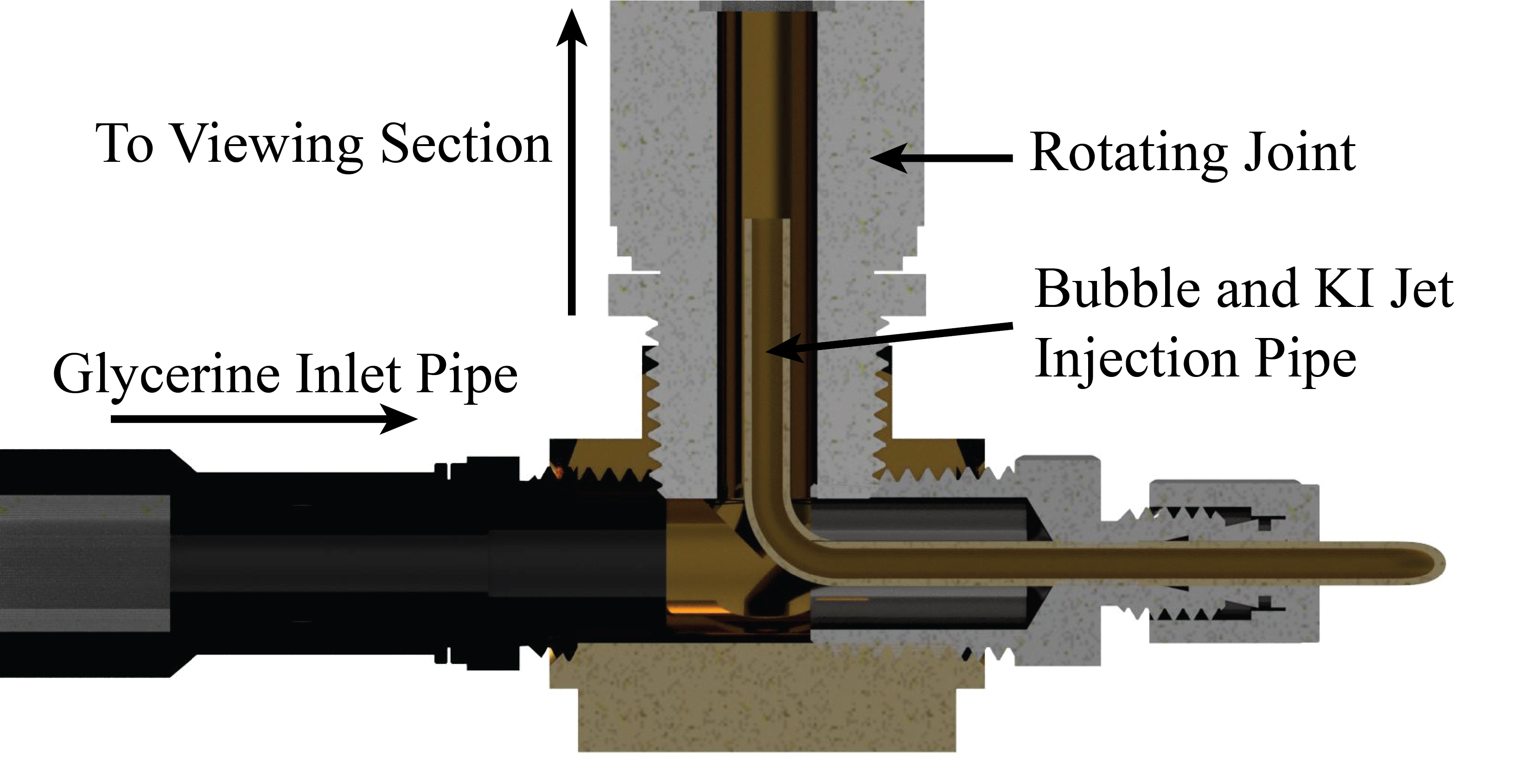}
         \caption{}
         \label{fig:flowsetup2}
     \end{subfigure}
     \caption{(a) Schematic of the flow setup where the test section is connected to supply lines via a swivel joint that rotates freely. (b) The cross-section of the flow setup shows a brass pipe terminating in the swivel joint is mounted concentrically with the plastic pipe. The Taylor bubble and KI jet are injected via this brass pipe.}
     \label{fig:flowsetup}
\end{figure}

Figure \ref{fig:flowsetup} shows the flow experiment setup. A 6.35~mm (0.25~in.) inner diameter, 9.53~mm (0.375~in.) outer diameter polycarbonate pipe is held at the top end to three linear motion stages (Optics Focus MOX-06-150, MOX-06-200, and MOX-06-100-B) and a LAB Motion Systems RT100S air bearing rotation stage. The RT100S air bearing stage has a maximum rotation speed of 3~Hz, which constrains the maximum the CT scan rate in these experiments. The pipe is filled with seeded liquid prior to the start of experiments. It is connected to a T-joint that enables a syringe pump to drive the flow or the injection of a secondary liquid or gas. The flow is driven by a Harvard Apparatus Pump 33 Dual Drive System (HAP33DDS). The HAP33DDS pump is able to drive two syringes independently and simultaneously. The first syringe contains the primary fluid (in this case glycerine); the other contains air for the Taylor bubble or KI-water solution for the scalar mixing jet experiments. The HAP33DDS is accurate to 0.25\% of the flow rate and can achieve flow rates ranging from 1.02~pL/min to 106~mL/min depending on the syringe size. Our experiments use glycerine and flows range from $Re = 8.7\times10^{-4}$ for the pipe flow TXPV experiments to $Re = 486$ for the KI jet.

A 3.18~mm (0.125~in.) diameter brass pipe is installed concentrically with respect to the plastic pipe, but terminates below the pipe in the T-joint as seen in figure \ref{fig:flowsetup2}. For the Taylor bubble experiments, this pipe is connected to the second syringe, which is full of air. The Taylor bubble is released from this pipe into the plastic pipe test section. For the KI jet experiments, this pipe is connected to a syringe full of KI-water solution, which is then ejected from the brass pipe into the test section.

For these experiments, we use pure glycerine as the primary fluid. The glycerine temperature is nominally constant at 27.6$\degree$C, the density is 1260~kg/m$^3$, and the dynamic viscosity is 0.73~Pa$\cdot$s. As a result of its higher viscosity, glycerine has an entrance length two orders of magnitude shorter than water would have had for the same velocity. This allowed us to test higher velocities while retaining a fully-developed pipe flow. Glycerine also retarded the rise of the Taylor bubble, making it possible to reconstruct its morphology with the 3~Hz full angle CT acquisition rate. The glycerine is seeded with hollow carbon tungsten-coated microsphere (CW) tracer particles developed and described in \cite{parker_enhanced_2022}.

\subsection{Experiment Parameters}
For the 2D-projected pipe flow experiments the glycerine-tracer mixture is pumped at 42.75~mL/min, yielding a center line speed of 45~mm/s ($Re = 0.40$). Images are captured at 1~kHz. A slower flow is used for the TXPV experiments to enable CT reconstruction with tolerable motion blur (further discussed in section \ref{ssec:motblur}). The pipe is rotated at 720 degrees per second (dps) for a full 360$\degree$ CT temporal resolution of 2~Hz. The fluid is pumped at 0.0760~mL/min, resulting in a 0.08~mm/s center line flow speed ($Re = 8.7\times10^{-4}$). At 720~dps, the centripetal acceleration is $0.05~g$ at the pipe wall, where $g$ is the acceleration due to gravity on earth, and is considered negligible.

Before image acquisition begins, the pipe is allowed to rotate for at least 5 seconds so that liquid will be experiencing solid body rotation when the data is captured. The characteristic time scale for flow inside an impulsively rotated cylinder is given by $\tau = R^2\nu^{-1}$ \cite{rivlin_spin-up_1983}, where $\nu$ is the kinematic viscosity of glycerine and $R$ is the pipe radius. At the temperature we operate at, $\tau = 0.02$~s -- two orders of magnitude less than the duration we wait before capturing data, and as expected the data do not show azimuthal velocity components.

In the Taylor bubble experiments 0.5~mL of air is injected through the brass ejection pipe within the T-joint. The CW-laden glycerine is already seeded into the viewing section. There is no pump-induced co-flow; the only flow is due to the Taylor bubble motion. The bubble rises at approximately 1.9~mm/s ($Re = 0.014$). Images are captured at a 1~kHz frame rate; the pipe is rotated at 1080~dps (3~Hz). Although for the Taylor bubble $\tau = 0.65$~s (assuming air fills the pipe), we do note a slight azimuthal component in our results. The average azimuthal velocity component is 0.1$\times$ the average axial component. At this rotation speed the centripetal acceleration in the pipe at the wall is $0.11~g$, which could begin to distort the Taylor bubble shape. As shown in section \ref{ssec:tracers}, though, this centripetal acceleration is not a major concern for the tracers since they are nominally density matched with the liquid. To increase the CT acquisition rate without concern for centripetal acceleration effects, one could rotate the source-detector pair as opposed to the flow loop (as done in \cite{dewanckele_innovations_2020}), or use multiple source-detector pairs.

In the KI jet experiments, the KI-water solution is nearly saturated at 1.4~g/mL of dissolved KI. We inject the solution at 37.21~mL/min. $Re = 486$ based on flow exiting the nozzle with speed of 78.1~mm/s, the density of the KI-water solution (1650~kg/m$^3$), the nozzle diameter, and the dynamic viscosity of water at 27.6$\degree$C. The jet is observed roughly 19 nozzle diameters downstream of the injection location. Images are captured at 500~Hz and the pipe is rotated at 720~dps (2~Hz). For finer temporal interrogation of the jet evolution, we reconstruct the CT scans by overlapping the projections used for each reconstruction by 90\%. While this does not increase the true temporal resolution -- which would mitigate motion artifacts discussed in section \ref{ssec:motblur} -- it better leverages the data contained in the 500 projections per scan to show the jet evolution in finer temporal detail. For simplicity, we measure only the evolution of the KI-water solution concentration in 3D -- velocimetry data is not captured in this experiment, but the concept is that simultaneous scalar field and flow field measurement would also be feasible. The parameters for all of the experiments can be found in table \ref{tab:expparam}.

\begin{table*}[]
    \centering
    \makebox[\textwidth][c]{
    \begin{tabular}{l c c c c c}
    Experiment & Fluid & $F$ (kHz) & $\omega$ (dps) & Re & $U_{max}$ (mm/s)\\
    \hline
    \hline
    2D Pipe Flow & Glycerine & 1 & 0 & 8.7$\times10^{-4}$ & 45 \\
    3D Pipe Flow & Glycerine & 1 & 720 & 0.40 & 0.08\\
    3D Taylor Bubble & Glycerine & 1 & 1080 & 0.014 & 0.5 \\
    3D KI Jet & Glycerine, KI-H\textsubscript{2}O & 0.5 & 720 & 486 & 78.1$^*$ \\
    \end{tabular}}
    \caption{The parameters for each of the three experiments that are conducted $F$ is the frame rate; $\omega$ is the pipe rotation rate; $U_{max}$  is the maximum expected flow speed from which the Reynolds number $Re$ is calculated. $^*$At the nozzle exit; we measure roughly 19 nozzle diameters downstream.}
    \label{tab:expparam}
\end{table*}

\subsection{Tracer Particles} \label{ssec:tracers}
Tracer particle properties determine how quantitative a particle-based velocimetry measurement can be. Ideally, tracer particles should be neutrally buoyant, small, and have low inertia such that they are faithful flow followers. Neutral buoyancy ensures that there are no bias errors in the vertical velocity component measurement or due to centripetal acceleration. Small particles maximize the spatial resolution of the measurement. Smaller particles can also perform better in presence of steep velocity gradients \cite{westerweel_velocity_2008}. Lastly, low inertia (for which being a small particle helps) means that the particle will more readily accelerate with the surrounding fluid. The parameter used to determine how well particles trace the flow is the Stokes number, $St$, defined by equation \ref{eq:stokesnum}. When $St \ll 1$, the particles can be said to be accurate flow tracers.
\begin{equation}\label{eq:stokesnum}
    St \equiv \frac{\tau_p}{\tau_f}
\end{equation}
In equation \ref{eq:stokesnum}, $\tau_p$ is the characteristic response time of the particle to an impulse uniform acceleration and $\tau_f$ is a characteristic time of the flow. Typically $\tau_f$ is taken to be a length scale divided by a characteristic velocity. For a particle that approaches neutral buoyancy, the response time can be calculated by
\begin{equation} \label{eq:taup}
    \tau_p = d_p^2 \frac{\Delta \rho}{18\mu}.
\end{equation}
where $d_p$ is the particle diameter, $\Delta \rho$ is the difference between the fluid and particle densities, and \textmu is the dynamic viscosity of the fluid. From equations \ref{eq:stokesnum} and \ref{eq:taup} it is clear that a particle that is small and neutrally buoyant (based on the St criteria, which also introduces a flow time scale dependence) is necessary for accurate measurement.

\subsubsection{CW Tracers}
Recently developed nominally neutrally buoyant 45--53~\textmu m tungsten-coated hollow carbon microsphere (CW) tracers are used in this study. The CW tracer particles, designed with tools developed in \cite{parker_experimentally_2022}, were shown in \cite{parker_enhanced_2022} to have higher contrast and better localizability than similarly sized silver-coated hollow glass tracer particles, and hence they are used for the high-speed measurements in these experiments. The CW particles are manufactured by Ultramet in California.

\begin{figure}
    \centering
    \includegraphics[width=0.5 \textwidth]{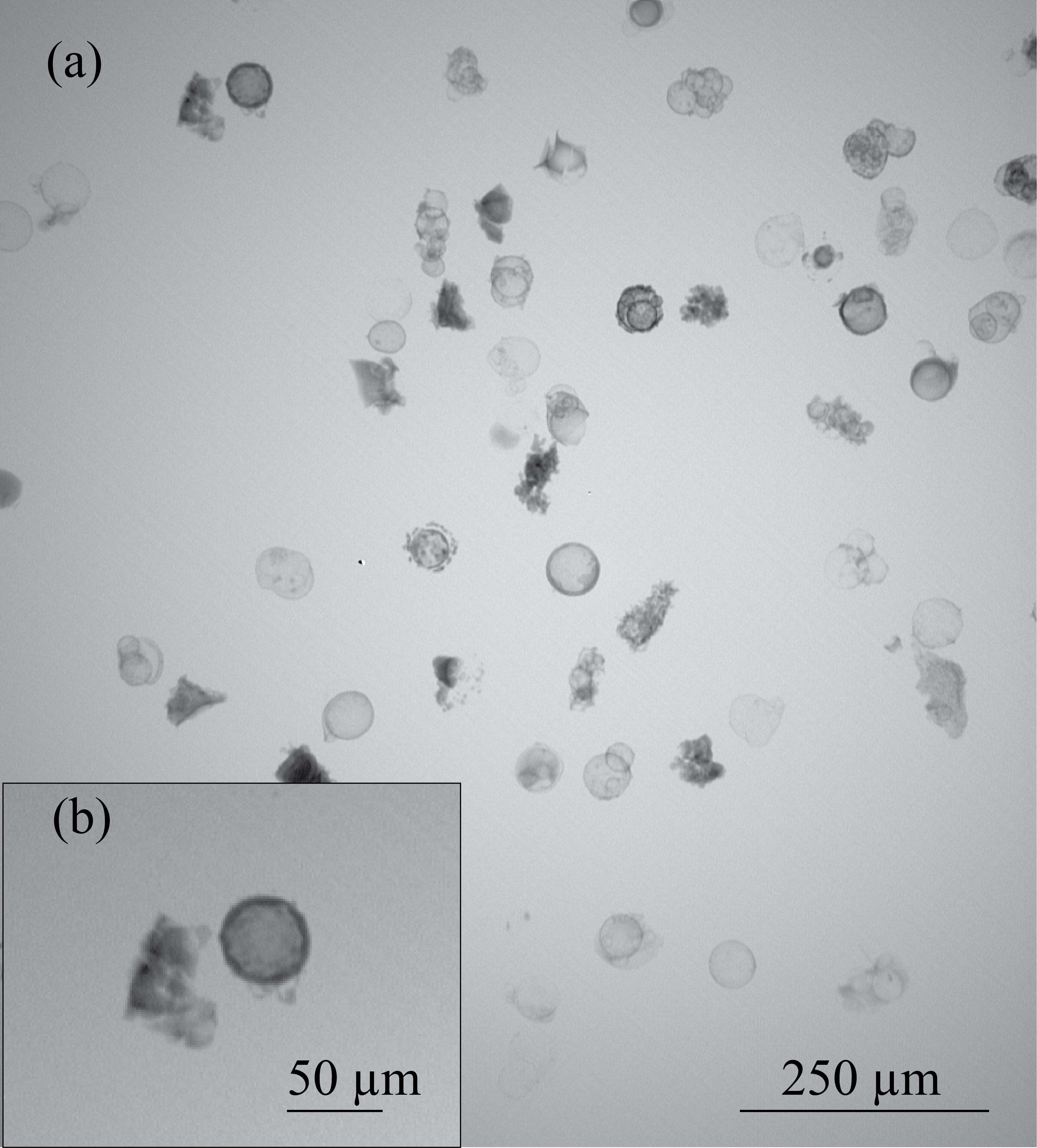}
    \caption{(a) Multiple batches, such as this one, of CW particles are examined with a nanofocus X-ray source (NanoTube N3, Excillum AB). (b) The prototype nature of these particles is evident from the various coating thicknesses, broken particles, and polydispersity.}
    \label{fig:nanocw}
\end{figure}

\begin{figure}
    \centering
    \includegraphics[width=0.65\textwidth]{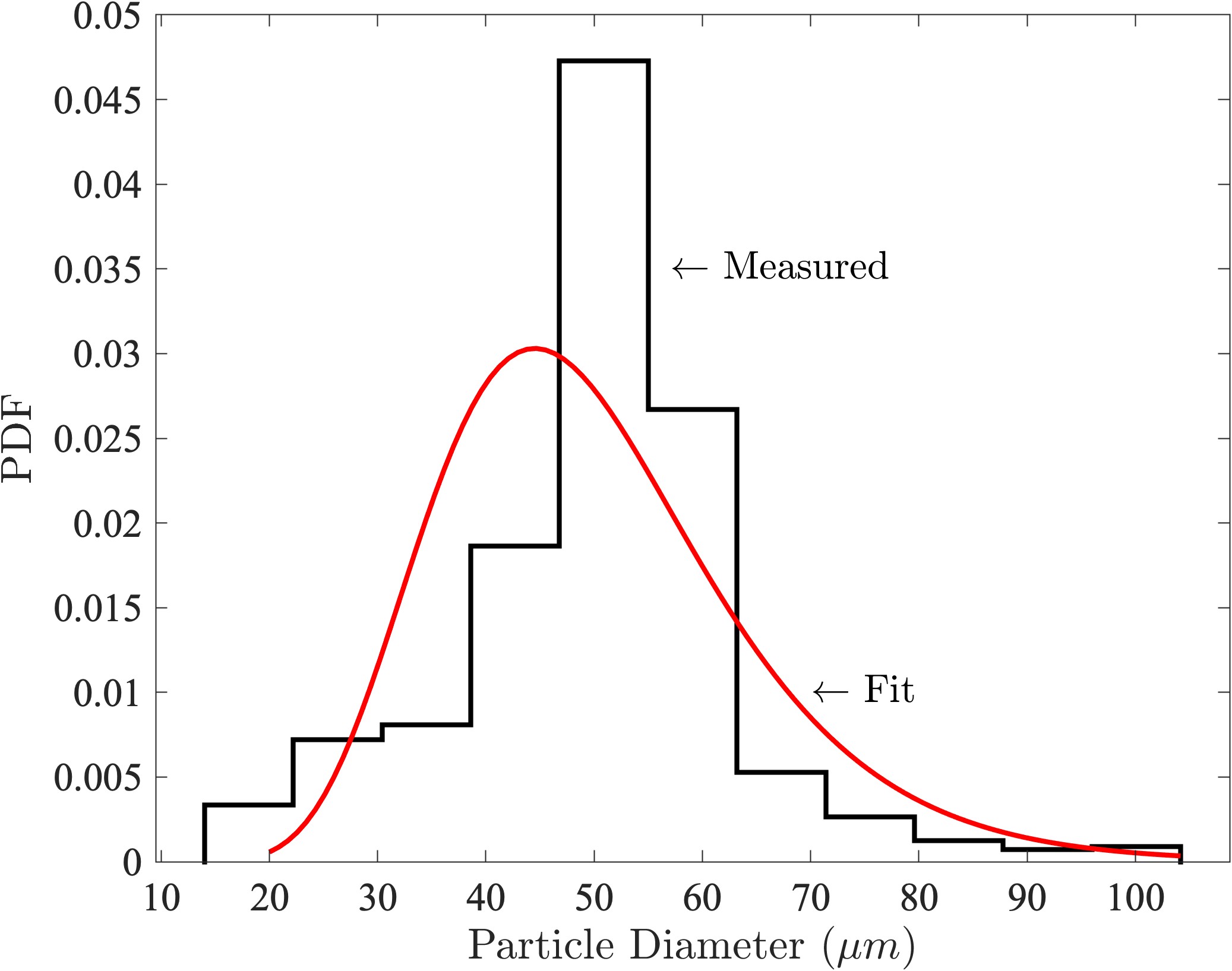}
    \caption{The measured and fit particle size distribution. We fit a log-normal distribution to the particles, which exhibit clear polydispersity.}
    \label{fig:cwsize}
\end{figure}

\begin{figure}
    \centering
    \includegraphics[width=0.65\textwidth]{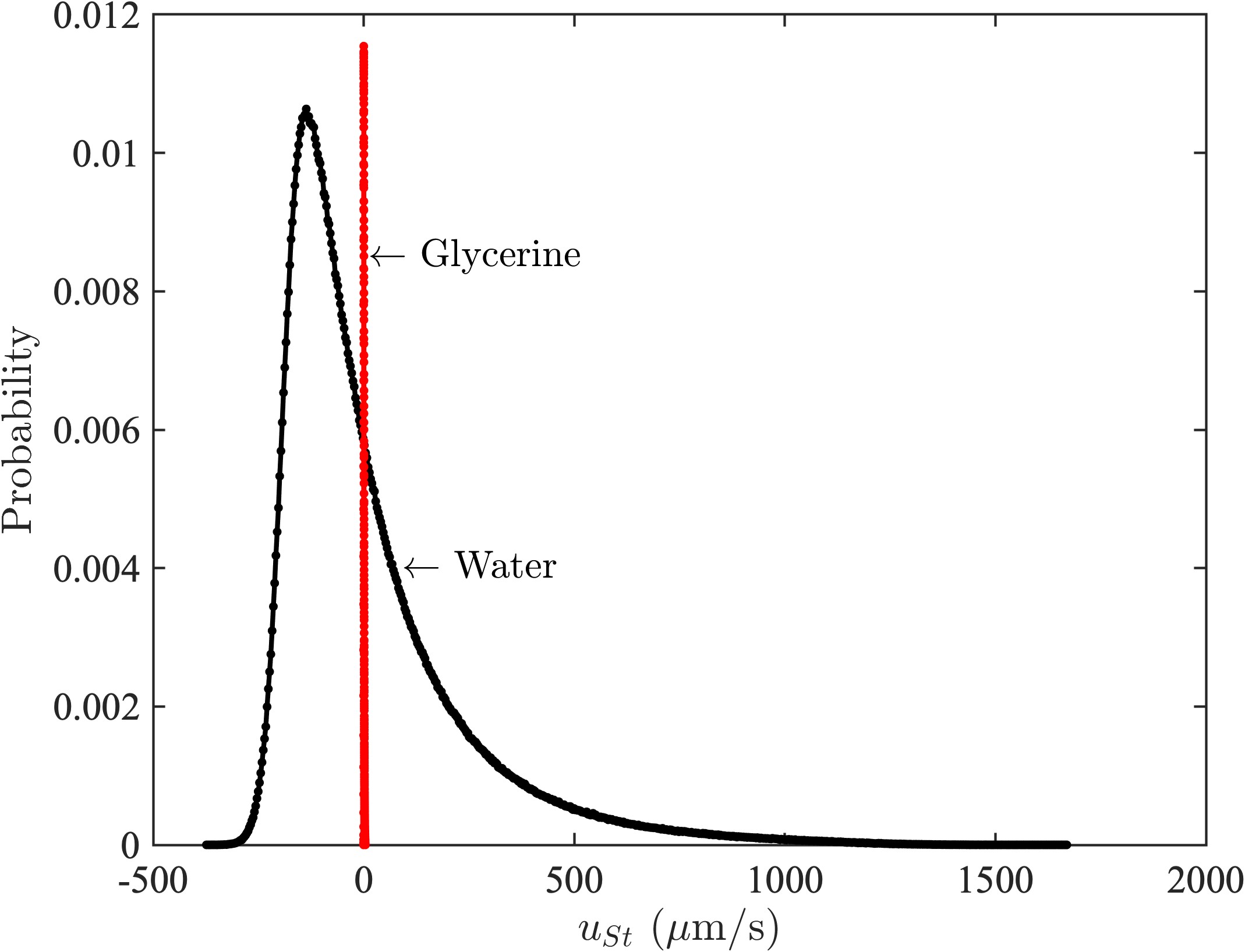}
    \caption{The settling speed distribution of the CW tracer particles in water and glycerine. Using glycerine dramatically reduces the spread in settling speeds compared to water and was hence used with these advanced prototype particles that lack the monodispersity mass production particles can be expected to achieve.}
    \label{fig:cwMC}
\end{figure}

We use a Monte Carlo calculation to estimate the settling speed distribution of the nominally monodisperce (but real and hence imperfect) tracer particles. Inherent to our calculations are the assumptions that 1) the tracer particles are spherical and 2) the relative velocity between the settling or rising particles is low enough that the particles are in the Stokes flow regime (i.e., with Reynolds number based on relative velocity to fluid less than unity). These particles are still in the prototype stage, however, so they exhibit strong polydispersity and uneven coating thicknesses. Based on SEM scan data we assume a 200~nm mean tungsten coating thickness that is Gaussian-distributed with a standard deviation of 20~nm. With this coating thickness and the manufacturer-reported uncoated particle density of 0.5767~g/cm$^3$ we calculate the carbon particle diameter range to be 40.4--47.6~\textmu m. Based on nanofocus X-ray images of the particles, we measure the particle diameter distribution used in the Monte Carlo calculations. An example nanofocus image can be seen in figure \ref{fig:nanocw}; the particle size distribution based on all such batches is shown in figure \ref{fig:cwsize}. We find the particles have an average diameter of 50~\textmu m, as designed. The standard deviation is large, as expected given the polydispersity, at 13~\textmu m.
In figure \ref{fig:cwMC} we show the settling velocity distribution of the CW particles in glycerine and water based on Monte Carlo calculations with $10^7$ trials. Table \ref{tab:particle_settle} lists the settling speed as a ratio of the characteristic flow speed and the Stokes number of CW particles in both water and glycerine. In glycerine, each of these values is much less than unity, so we can take the CW particles to be ideal flow tracers.
\begin{table}[]
    \centering
    \begin{tabular}{ l c c }
        Fluid &  $\lvert \left< u \right>_{St} \rvert / U$  &  $St \times 10^6$   \\
        \hline \hline
        Glycerine & 6.8$\times 10^{-4}$ & 2.8 \\
        Water & 0.019 & 2300\\
    \end{tabular}
    \caption{The mean settling speed to center line speed ($U$) ratio and Stokes number of the CW tracer particles that we use in these experiments. For the speed ratio, we compare to the slowest flow speed we measured ($U=0.08$~mm/s); for the Stokes number we use the highest flow speed of $U=45$~mm/s. This is to check if our Stokes number and buoyancy measurement bias can be expected to be negligible for all of the considered flow conditions.}
    \label{tab:particle_settle}
\end{table}

\subsection{The MetalJet X-ray Source}
\begin{figure}
     \centering
     \includegraphics[width=0.65\textwidth]{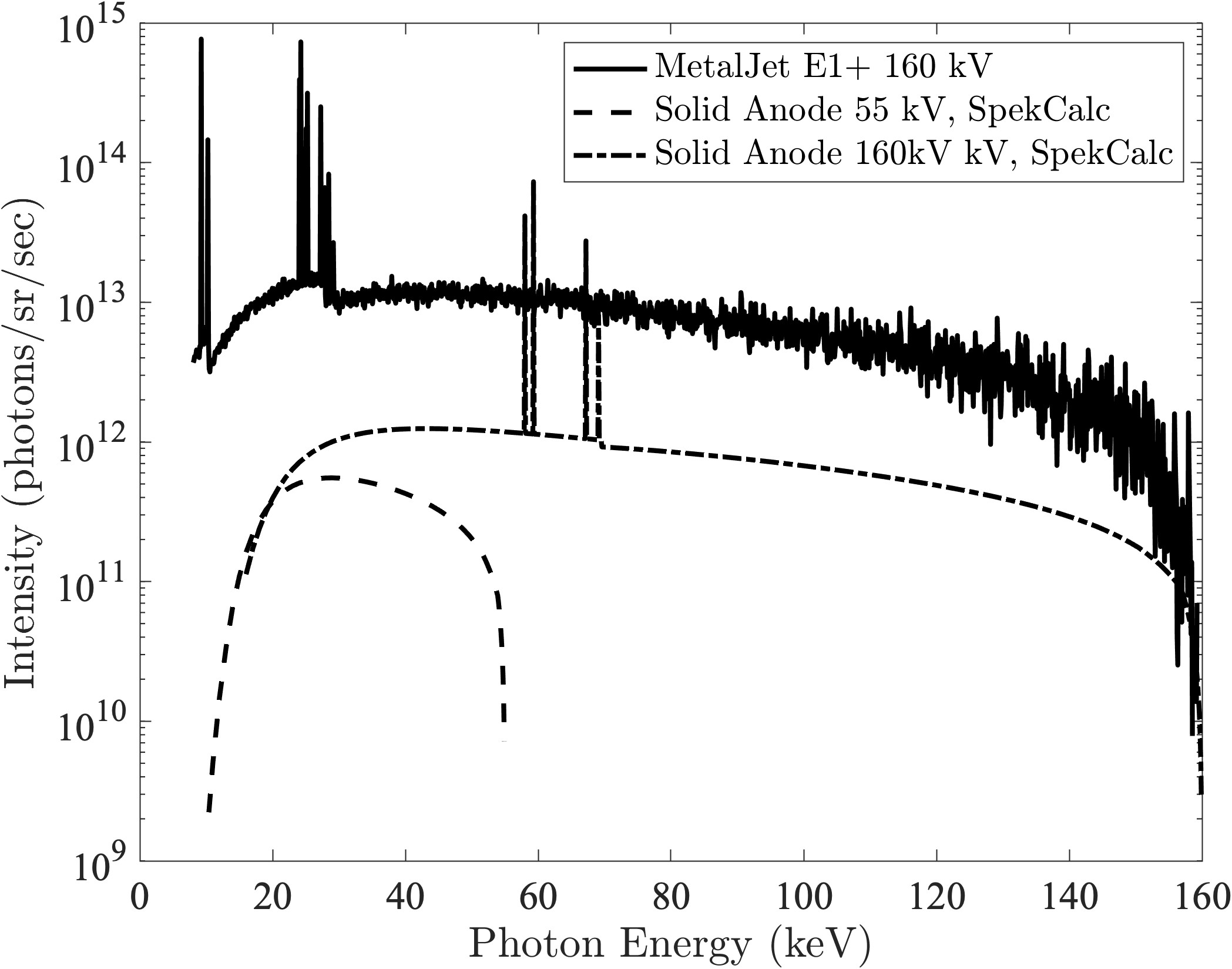}
     \caption{The X-ray flux from the MetalJet E1+ LMJ source used (\cite{excillum_metaljet_2023}) versus a traditional solid anode microfocus X-ray source \cite{poludniowski_spekcalc_2009, poludniowski_calculation_2007, poludniowski_calculation_2007-1}. A LMJ is able to withstand greater focal spot power density. As a result the spectrum is 20$\times$ brighter for the same focal spot size and source acceleration voltage.}
     \label{fig:spectra}
\end{figure}

A critical element that, combined with the enhanced particles and detector, enables \textit{O}(1~kHz) frame rates is the MetalJet E1+ X-ray source from \cite{excillum_metaljet_2023}. This source is an order of magnitude brighter than a traditional solid anode microfocus X-ray source. For example, the MetalJet E1+ model used is roughly 20$\times$ brighter for the same focal spot size and source acceleration voltage than the solid anode source used in \cite{parker_enhanced_2022}. The apparent X-ray focal spot size is approximately 30~\textmu m at the settings used in this study. We utilize a 160~kV source acceleration voltage and 4.375~mA target current (700~W) compared to 55~kV and 0.5~mA (25.5~W) for the source in \cite{parker_enhanced_2022}, resulting in a nominally 83$\times$ brighter source compared to that study. When matching the focal spot size and source acceleration voltage, the MetalJet E1+ source is 20$\times$ brighter than the YXLON source in \cite{parker_enhanced_2022}. The spectra for the MetalJet E1+ and a traditional solid anode microfocus tube used in \cite{parker_experimentally_2022} and \cite{parker_enhanced_2022} can be seen in figures \ref{fig:spectra} for a rough comparison of a LMJ source to a typical solid anode microfocus source. Here, brightness is defined as the total number of photons emitted per steradian per second. While convenient, this definition does not account for different photon energy spectra between sources. Different spectra may affect the image quality improvement depending on the materials in the field of view and detector.

\subsection{Image Processing}
We collect data images with XC-Thor FX20.1.256 photon counting detector (PCD), with the energy threshold 15~keV. For this study the flat field corrections are applied by the detector software prior to any custom processing of data. The flat field correction map is generated by the detector software with images taken after placing 0, 2, 7, and 21~mm thick plates of aluminum in front of the source one-by-one and collecting flat images at the source settings to be used \cite{davidson_limitations_2003}.

For 2D-projected XPV processing, the flat field corrected images are inverted, then imported to LaVision DaVis 8.4. We use standard DaVis algorithms as detailed in the appendix. For TXPV, we reconstruct the field of view using the ASTRA Toolbox \cite{van_aarle_astra_2015, aarle_fast_2016}. Then, we segment the particles from the fluid in MATLAB to extract the particle centroid locations. A particle tracking code in MATLAB was developed to trace the particle movement through multiple reconstructions. Workflows for the pipe flow, Taylor bubble flow, and KI jet flow are detailed in appendices.

\subsubsection{Calibration and Uncertainty for 2D XPV}
We define the pixel-to-mm calibration by counting the pixels required to span the known pipe outer diameter -- 9.5~mm (0.375~in.). As discussed in \cite{parker_enhanced_2022}, the actual pixel-to-mm scaling changes as a function of particle distance from the source (within the pipe) due to geometric magnification. However, because we collect 2D-projected images, this depth information is lost, so a single calibration value is used for the entire experiment depth. A useful rule of thumb is to keep the relative error associated with geometric magnification, termed the relative magnification error (RME), below 10\%. \cite{parker_enhanced_2022} derives the equation for the RME, shown to be
\begin{equation} \label{eq:magerror}
    \varepsilon (y) = \frac{ \lvert \Delta z - \Delta Z \rvert }{ \Delta Z} 
    = 1 - \left(1 + \frac{\lvert y \rvert}{SOD} \right)^{-1},
\end{equation}
where $\Delta z$ and $\Delta Z$ are the measured tracer particle motion and the true particle motion, respectively; $y$ is the depth-wise location of the tracer particle measured from the center of the experiment. In order to keep the maximum RME below 10\%, the magnification aspect ratio ($MAR$) should be kept below 0.2. The $MAR$ is defined in equation \ref{eq:mar}, where $\delta$ is the maximum flow domain depth (pipe diameter in present study) in the beam direction.
\begin{equation}\label{eq:mar}
    MAR = \frac{ \delta }{ SOD }
\end{equation}
Equation \ref{eq:mar} assumes symmetry in the depth-wise direction ($\delta = 2~y_{max}$). For all $y < y_{max}$, the RME will then be less than 10\%. For these experiments, the $MAR = 0.11$.

Another source of measurement variance that is unique to 2D-projected data is the depth-averaged velocity profile (DAVP). In 2D-projected data, particles at the back wall, center, and front wall can all appear at the same horizontal location across the pipe. As a result, when collecting particle tracking data, one is actually sampling from the velocity distribution in the depth-wise direction at a given horizontal location. In flows with strong velocity gradients, this can be problematic. Even in flows with relatively shallow gradients, such as laminar pipe flow, the measured velocity profile is depressed compared to the radial cross section velocity profile. For laminar pipe flow, where there is an analytical solution, the DAVP can be calculated by taking a depth-wise average, shown in equation \ref{eq:davp}.
\begin{equation} \label{eq:davp}
    \begin{split}
\left<u(x,y)\right>_y &= \frac{4 Q}{3 \pi R^2} \left[1 - \left(\frac{x}{R}\right)^2\right]
= \frac{2}{3} u(r=x)
\end{split}
\end{equation}
It is to this profile that we compare our data. Here, $\left<\cdot\right>_i$ denotes mean in the $i$ direction, $Q$ is the volumetric flow rate, $R$ is the pipe radius, and $x$ is the horizontal coordinate (aligned here with the projected radial coordinate). Measuring from a distribution of velocities that ranges from $u=0$ to center line speed, $U$, means that there will inevitably be a wide spread in the velocity measurement. As with the RME, issues with depth-averaged profiles can be alleviated with 3D tomographic or stereo XPV measurements.

The final source of uncertainty considered is buoyancy, which is discussed in section \ref{ssec:tracers}. Buoyancy effects in vertically aligned flows can create a velocity measurement bias. We showed in section \ref{ssec:tracers} that the flow speeds are sufficiently high compared to the particle settling speeds that buoyancy error can be neglected.

\subsubsection{Calibration and Uncertainty for TXPV}
For TXPV we define a voxel-to-mm$^3$ calibration similar to the 2D calibration. We identify the outer diameter of the pipe -- known to be 9.5~mm -- and count the number of pixels required to traverse it. Since the pixels are square, the reconstruction voxel volume is set to the pixel width cubed.

TXPV is subject to sources of uncertainty different from those encountered in visible light tomographic PTV. As in visible light, TPIV particles are segmented in a reconstructed volume; the particle volume centroid is taken to be its location. Typically, in visible light PIV multiple cameras capture an instantaneous snapshot with minimal blur due to a laser pulse duration \textit{O}(100~ns). By contrast, in TXPV the tracer particles and flow morphology (e.g. Taylor bubble) often are moving appreciably during the CT scan. This can result in a number of artifacts that may smear or warp the particle and bubble reconstruction, some of which were illustrated in appendix of \cite{makiharju_tomographic_2022}. In order for the particles and bubble to appear well defined in the reconstruction, it is crucial that the scan time is faster than the particle and morphology motion. ``Faster" as a rule of thumb can be taken to mean that a particle does not move more than one diameter during the scan, although in some circumstances two diameters can be acceptable. On the other hand, if the rotation rate is high particles or bubbles may be blurred by rotational motion in the individual projections if the exposure time is too high. Balancing these effects is important for getting a usable CT scan for TXPV. Section \ref{ssec:motblur} discusses this balance in greater detail.

Table \ref{tab:motionblur} shows the expected maximum particle rotation blur and particle motion blur (projection or tomographic) during scanning. These blurring phenomena create motion artifacts in the reconstruction. We use the maximum expected velocities, the rotation speed, and the maximum particle radial distance to calculate the values in table \ref{tab:motionblur}, so these values are conservative. Based on the values in table \ref{tab:motionblur}, we would expect to see some blurring and artifacts. For the present experiments the motion artifacts are sufficiently modest to enable particle tracking.

\begin{table}[]
    \centering
    \begin{tabular}{l c c}
    Experiment & Particle Rotation Blur & Particle Motion Blur \\
    \hline
    \hline
    3D Pipe Flow & 0.8$d_p$ & 0.8$d_p$ \\
    3D Taylor Bubble & 1.2$d_p$ & 3.33$d_p$ \\
    \end{tabular}
    \caption{The expected maximum particle blur due to pipe rotation and particle motion in units of nominal particle diameter, $d_p$ (50~\textmu m). Maximum expected flow velocities and radial distances are used to calculate these values. Larger values indicate more blurring that can be expected. Beyond two particle diameters of motion the ability to localize the particle can deteriorate.}
    \label{tab:motionblur}
\end{table}

\begin{figure}
    \centering
    \includegraphics[width = 0.6\textwidth]{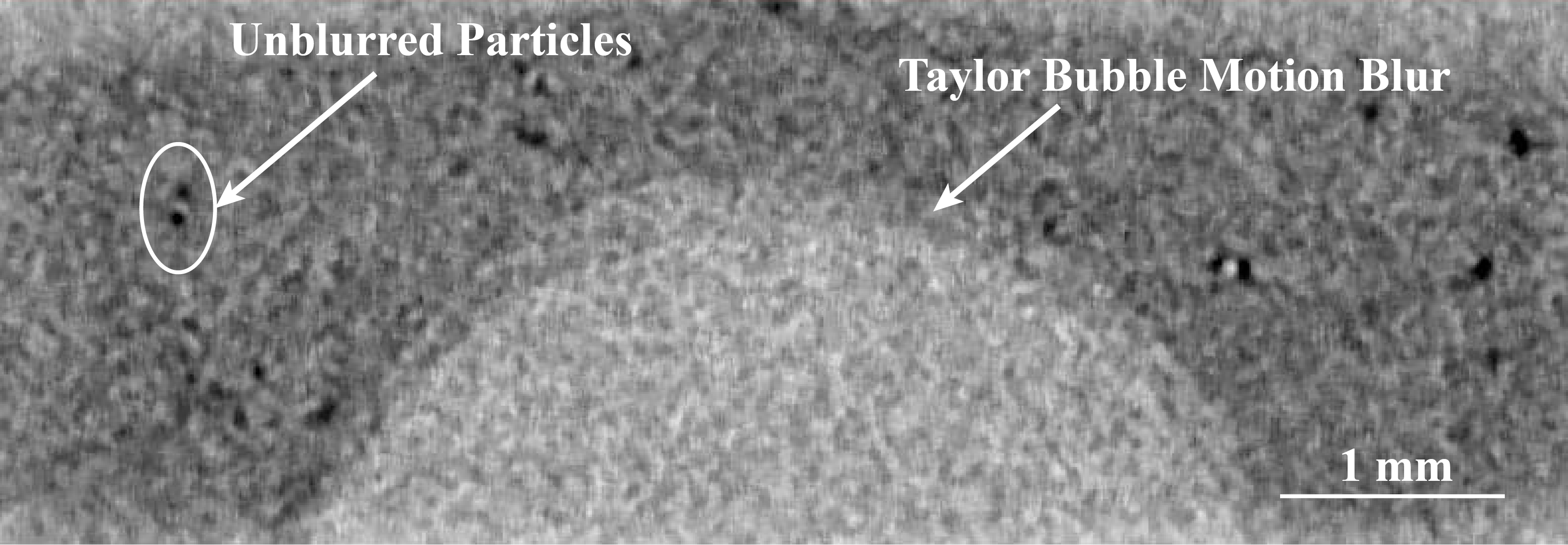}
    \caption{A $5 \times 5 \times 5$ median-filtered vertical slice through the central plane from a reconstructed Taylor bubble experiment. Only the pipe inner diameter is shown here for clarity. The Taylor bubble exhibits some motion blur manifested as a roughly 200~\textmu m thick ``halo" around the bubble, while in reality the interface is sharp. Fast moving particles may similarly warp, but many particles do not. There is a sufficient number of unwarped particles to successfully track them.}
    \label{fig:recon_artifacts}
\end{figure}

Figure \ref{fig:recon_artifacts} shows a vertical cross section through a Taylor bubble flow reconstruction. The bubble itself exhibits a motion blur artifact because it is rising too quickly for our scan speed. Although some particles may appear blurred, there are numerous particles that do not exhibit blurring and warping in the reconstruction, making tracking feasible.

In order to mitigate the error introduced by particle distortions (appendix of \cite{makiharju_tomographic_2022}), we introduced a radius match to the particle tracking cost function in order to make sure particle volumes of the same size are tracked as one particle. Particles that experience a large change in reconstructed volume are not tracked, mitigating error introduced from particle warping.

\section{Results} \label{sec:resdic}

\subsection{Poiseuille Pipe Flow}
\subsubsection{2D-projected Measurements of Poiseuille Pipe Flow}
\begin{figure}
    \centering
    \includegraphics[width=0.45\textwidth]{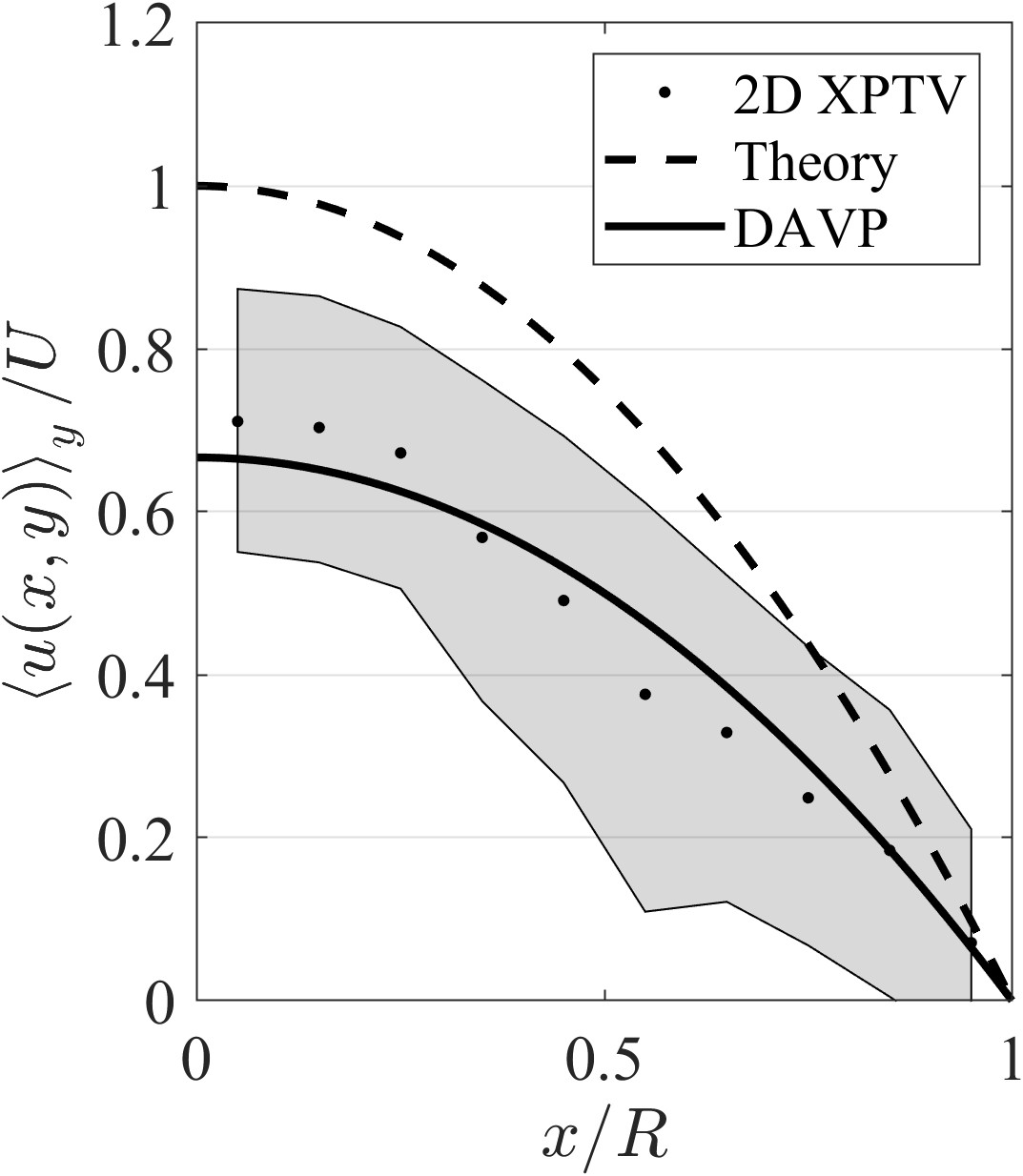}
    \caption{The depth-averaged velocity profile of a pipe flow with a 45~mm/s center line speed ($Re = 0.40$) agrees well with the expected profile. The gray region depicts the variation of the velocity measurement between the 16$^{th}$ and 84$^{th}$ percentiles, which are analogous to the 2$\sigma$ error bounds for a non-Gaussian-distributed variable. The data are normalized by the theoretical center line speed.}
    \label{fig:CW_DAVP}
\end{figure}

Figure \ref{fig:CW_DAVP} shows the DAVP as measured by CW tracer particles in a 45~mm/s center line speed flow ($Re = 0.40$). As shown, the data agree reasonably well with the analytical solution for the depth-averaged velocity profile. This profile is averaged over 850 frames to gather converged particle tracking statistics and assess the 2D-projected variance.

While these 2D measurements from a single source-detector pair are of limited value, multiple source-detector pairs can triangulate the location of particles in 3D in what is known as stereo-XPV (SXPV). SXPV would not require the rotation of source-detector pairs or the flow experiment and the temporal resolution would only be limited by the detector frame rate. In other words, 1~kHz SXPV is possible if multiple sources and detectors are used.

\subsubsection{TXPV of Poiseuille Pipe Flow}
\begin{figure}
    \centering
    \includegraphics[width=0.45\textwidth]{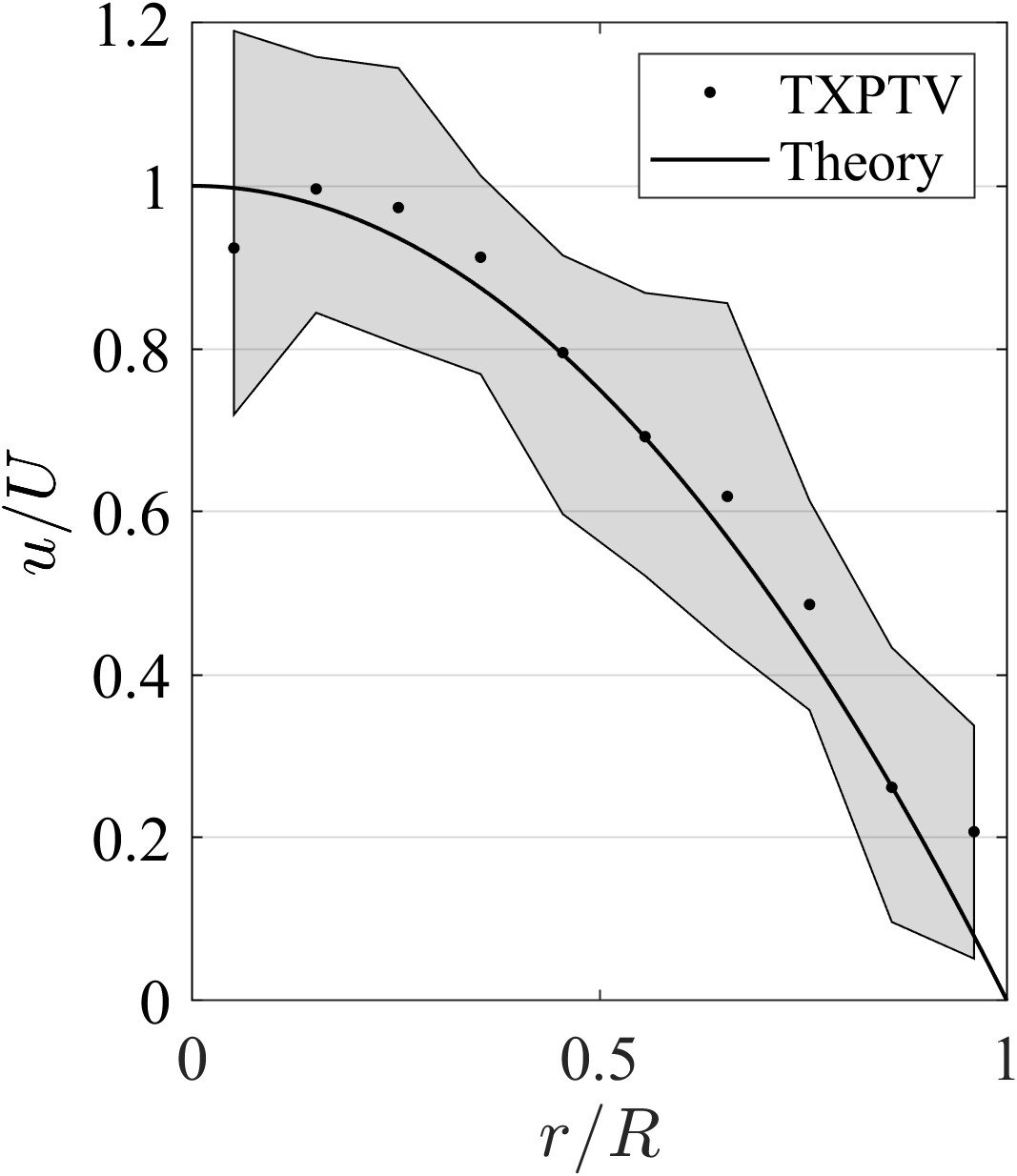}
    \caption{The TXPV radial flow profile from a single scan shows good agreement with the theoretical Poiseuille profile. The 16$^{th}$ and 84$^{th}$ percentiles are shown to approximate 2$\sigma$ error bounds for a non-Gaussian-distributed variable. Unlike 2D-projected XPV, TXPV can measure the true radial flow profile. The data are normalized by the theoretical center line speed, $U = 0.08$~mm/s.}
    \label{fig:3Dpipeprof}
\end{figure}

Figure \ref{fig:3Dpipeprof} shows the average velocity profile from six reconstructions captured at 2~Hz. This rotation speed is 4.5$\times$ faster than previously demonstrated in \cite{makiharju_tomographic_2022}. The measured profile shows good agreement with the theoretical Poiseuille pipe flow profile, demonstrating that TXPV can accurately measure fluid flow at 2~Hz in \textit{O}(cm) domain. Although Poiseuille pipe flow is a simple, well-understood flow, it can serve as a useful benchmark for steadily improving XPV measurements.

The relatively large spread in the data is likely due to particle motion artifacts in the reconstruction. \cite{makiharju_tomographic_2022} explains in detail the root cause of these artifacts. The relative rotation between the moving particles and the source-detector pair smears the particles in the reconstruction, making it difficult to locate their true centroids. Reducing rotation time relative to the flow velocity or employing limited angle reconstruction can reduce the impact of these artifacts. However, in the present study we focus less on the possibilities for data post processing, as these are the topic of a separate study that is in preparation.

\subsection{TXPV of Flow Around a Rising Taylor Bubble}
\begin{figure}
    \centering
    \includegraphics[width=0.6\textwidth]{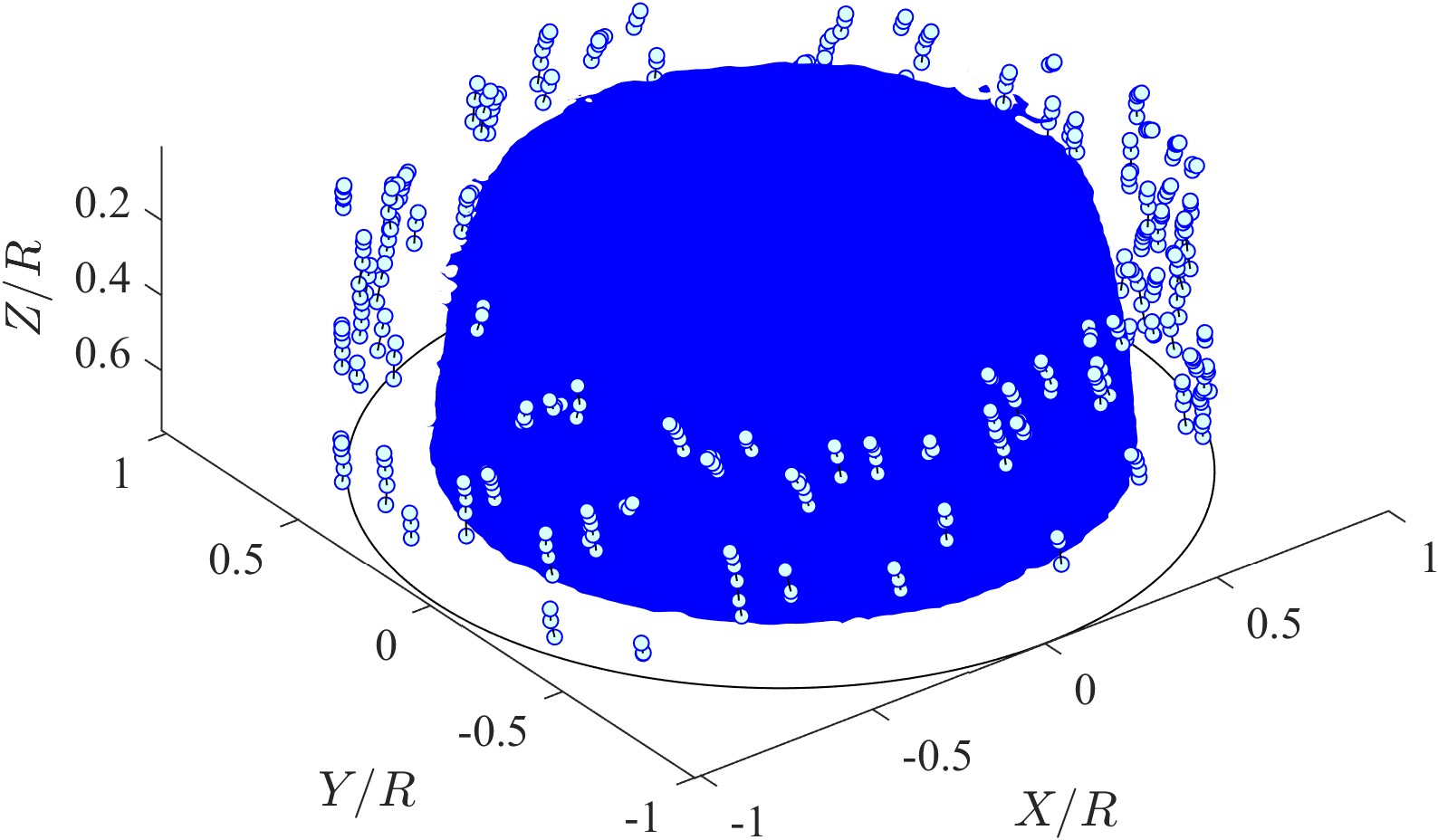}
    \caption{The particle pathlines flowing around the Taylor bubble moving through the field of view. The bubble-glycerine interface is shown in blue. A black circle on the bottom indicates the theoretical pipe interface.}
    \label{fig:taylor_vector}
\end{figure}

One benefit of fast TXPV is that it becomes possible to reconstruct the flow morphology in addition to capturing the flow velocity. For example, a full CT reconstruction of the Taylor bubble can be seen in figure \ref{fig:taylor_vector} with the pathlines of tracked particles flowing around it. These data are captured at 3~Hz, which is 9$\times$ faster than previously achieved by \cite{makiharju_tomographic_2022}. Furthermore, due to the higher frame rate we use here, we can capture 333 frames per rotation as opposed to the 200 frames per rotation used in \cite{makiharju_tomographic_2022}. More frames improves reconstruction spatial resolution and reduces noise, making particle detection more accurate and ultimately enabling the use of smaller particles.

A near order of magnitude improvement in the temporal resolution over the previous state of the art exemplifies the rapid development of TXPV specifically and XPV in general. As new, brighter X-ray sources become available alongside faster detectors TXPV will become increasingly useful as a technique for studying optically opaque fluid flows.

\subsection{Scalar Mixing of a KI Jet}
\begin{figure}
    \centering
    \includegraphics[width=0.6\textwidth]{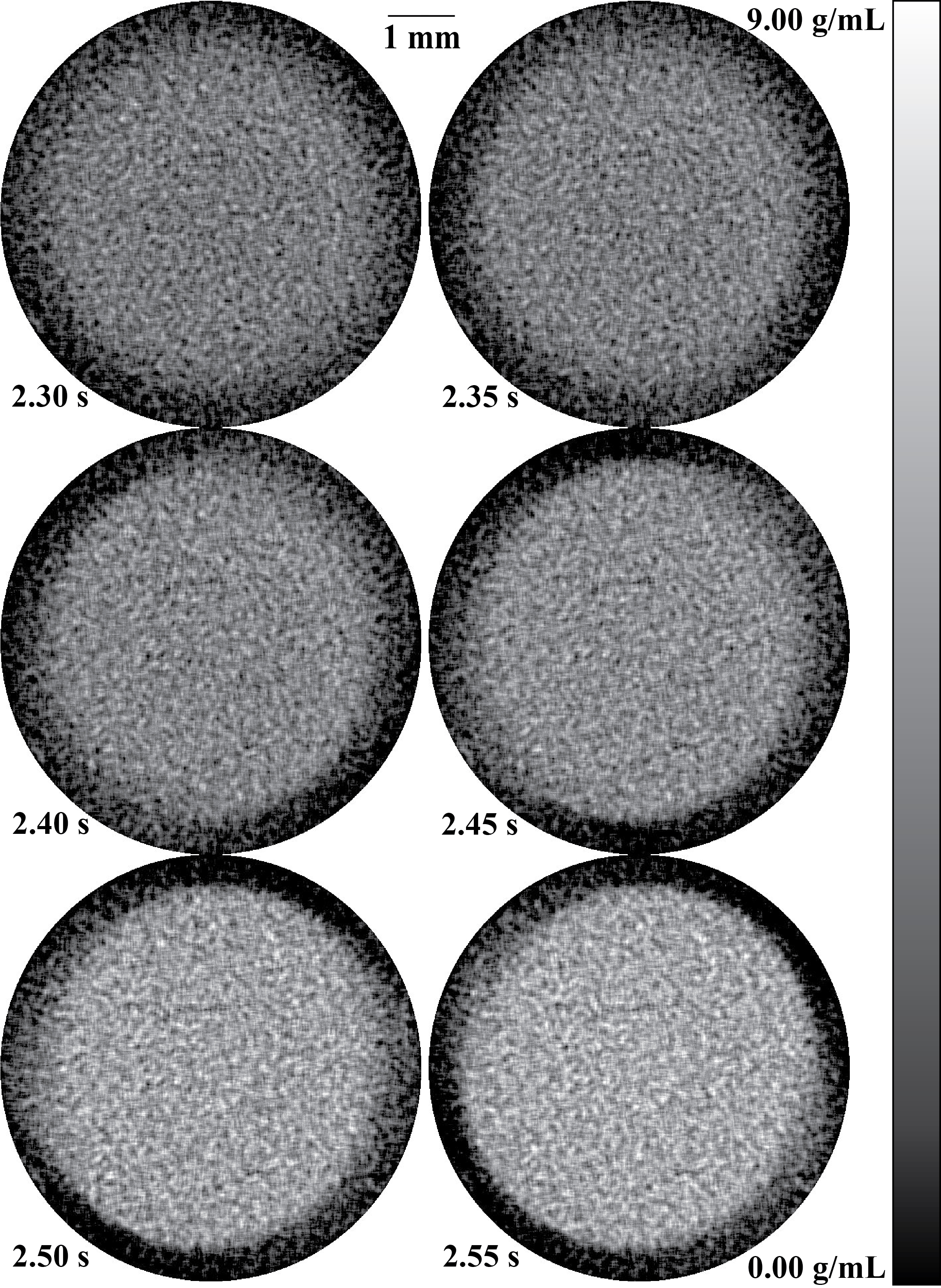}
    \caption{The evolution of the jet, but with the intensity values converted to KI-water solution concentration. The pipe wall is masked out for clarity. The reconstructed slices are Gaussian filtered with an $11 \times 11 \times 11$ kernel and a 7 pixel standard deviation.}
    \label{fig:conc_evol}
\end{figure}

Figure \ref{fig:conc_evol} shows the evolution of the KI jet as it passes through the center slice of the reconstructed volume. The KI-water solution attenuates X-rays more than glycerine, so the KI jet appears darker in raw images but lighter in the reconstruction with the chosen color scale. Initially, a more diffuse solution becomes visible; after 0.25~s the jet is more sharply visible. We can calculate the concentration of KI-water solution in glycerine by using the voxel intensity. According to the Beer-Lambert law for a mixture,
\begin{equation}\label{eq:blmix}
    N = N_0 \exp \left(-\sum \sigma_i n_i(c) x_i \right)
\end{equation}
the number of detected photons, $N$, is a function of the number of source photons $N_0$, the attenuation cross section of material $i$, $\sigma_i$, the atomic number density $n_i$ as a function of concentration $c_i$, and the material thickness $x_i$. For a given material -- in this case KI-water mixture -- the attenuation cross-section and material thickness (i.e., voxel size) are constant. The source intensity is assumed to be effectively constant. Flickering, from either the source or the detector, are accounted for with a correction factor in post-processing. Thus, the number of detected photons for a given voxel is a function only of the concentration in said voxel. Atomic number density is proportional to concentration, so we can calculate from equation \ref{eq:blmix} the concentration 
\begin{equation}\label{eq:concen}
    c = A \ln(N) + B,
\end{equation}
where $A$ and $B$ are constants that are determined by examining the voxel intensity with known concentrations. We can identify the voxel intensity at zero concentration by assuming no solution is in the first four reconstructions. Then, by assuming the solution is well mixed in the four final reconstructions, we can identify the intensity at the well mixed concentration. The well mixed concentration, based on a control volume of 1.9~mL and the injected volume in 4~sec of 2.48~mL, is 2.15~g/mL. The control volume here is taken to be the pipe length from the nozzle injection point to the viewing section of the pipe. Data is collected over 4~sec. We find the coefficients $A=-116$ and $B = 1216$. Figure \ref{fig:conc_evol} shows the concentration of the KI-water solution in the reconstruction as a function of time. As the jet develops in the field of view, the concentration of the solution increases. The qualitative data in each reconstruction is transformed into quantitative data in figure \ref{fig:conc_evol}. This demonstrates the potential of high speed CT scans for measuring the time evolution of scalar mixing in 3D. Although not done here, these data could be combined with TXPV measurements to simultaneously capture the evolution of fluid velocity and concentration fields.

\section{Designing In-Lab XPV Experiments}\label{sec:design}
Special considerations are necessary for designing in-lab XPV. Compared to synchrotrons geometric magnification is usually obtained with little effort, but the experiment is likely photon starved. That is, one often needs to design the flow loop, tracer particles, and choose imaging parameters such that usable image quality is achieved. For in-lab XPV with a PCD, XPV-specific tracer particles, and a LMJ source is a promising hardware combination going forward, particularly as these components improve. High-speed 2D in-lab X-ray imaging is a critical enabler for quantitative XPV. The frame rates and particle tracking achieved in this study are unique to these experiments, though. X-ray image quality is highly dependent on the experiment materials, flow media, experiment size, source spectrum, and more. In this section we will discuss a procedure to assess the feasibility and design of high speed in-lab XPV systems.

\subsection{Contrast}
Typically, XPV is contrast-limited. It is challenging to have particles that are neutrally buoyant, but with X-ray attenuation drastically different from surrounding fluid; small enough to be good flow followers, but large enough to generate significant contrast and be resolved in a large FOV. Presently the set of flows that can be examined with XPV is limited by the contrast that is achievable with currently available in-lab X-ray imaging capabilities. The experiments above have sought to advance these capabilities. To design a different XPV experiment the first step is identifying the feasibility XPV by estimating the maximum achievable contrast.

Let us define the contrast, $C$, for a PCD as
\begin{equation} \label{eq:contrast}
    C = \mid N_{bg} - N \mid
\end{equation}
where $N$ is the number of detected photons at the projected location of an object of interest and $N_{bg}$ is the number of photons detected in the background immediately surrounding said object. Note that this is attenuation-based contrast. Phase contrast imaging can also be used, but is not discussed here.

Ignoring scatter from domain or within detector, fluorescence, and other complicating -- but often second order -- factors the simplest way to estimate the contrast is the Beer-Lambert law, shown in equation \ref{eq:BLlaw}. Equation \ref{eq:blmix} is a simplified version of equation \ref{eq:BLlaw}, which assumes a spatially uniform source emission.
\begin{equation} \label{eq:BLlaw}
    N = \int_{E_{thresh}}^{E_{max}} I_0'(\varepsilon) \eta(\varepsilon) A t_{exp} \exp \left[ -\int_{\underline{x} \in V} \rho(\underline{x}) \mu(\underline{x}, \varepsilon) x d\underline{x} \right] d\varepsilon
\end{equation}
Here, $N$ is the number of photons, $\eta$ is the detector quantum efficiency (the ratio of detected photons to arriving photons), $E_{thresh}$ is the detector threshold energy, $E_{max}$ is the maximum photon energy produced by the X-ray source, $I_0'(\varepsilon)$ is the rate of photon emission from the X-ray source per unit time as a function of photon energy $\varepsilon$, $A$ is the detector pixel area, $t_{exp}$ is the exposure time, $V$ is the field of view (FOV) volume, $\rho(\underline{x})$ is the density, $\mu(\underline{x},\varepsilon)$ is the mass attenuation coefficient, and $\underline{x}$ is the location in the FOV volume.

\begin{figure}
    \centering
    \includegraphics[width=0.5\textwidth]{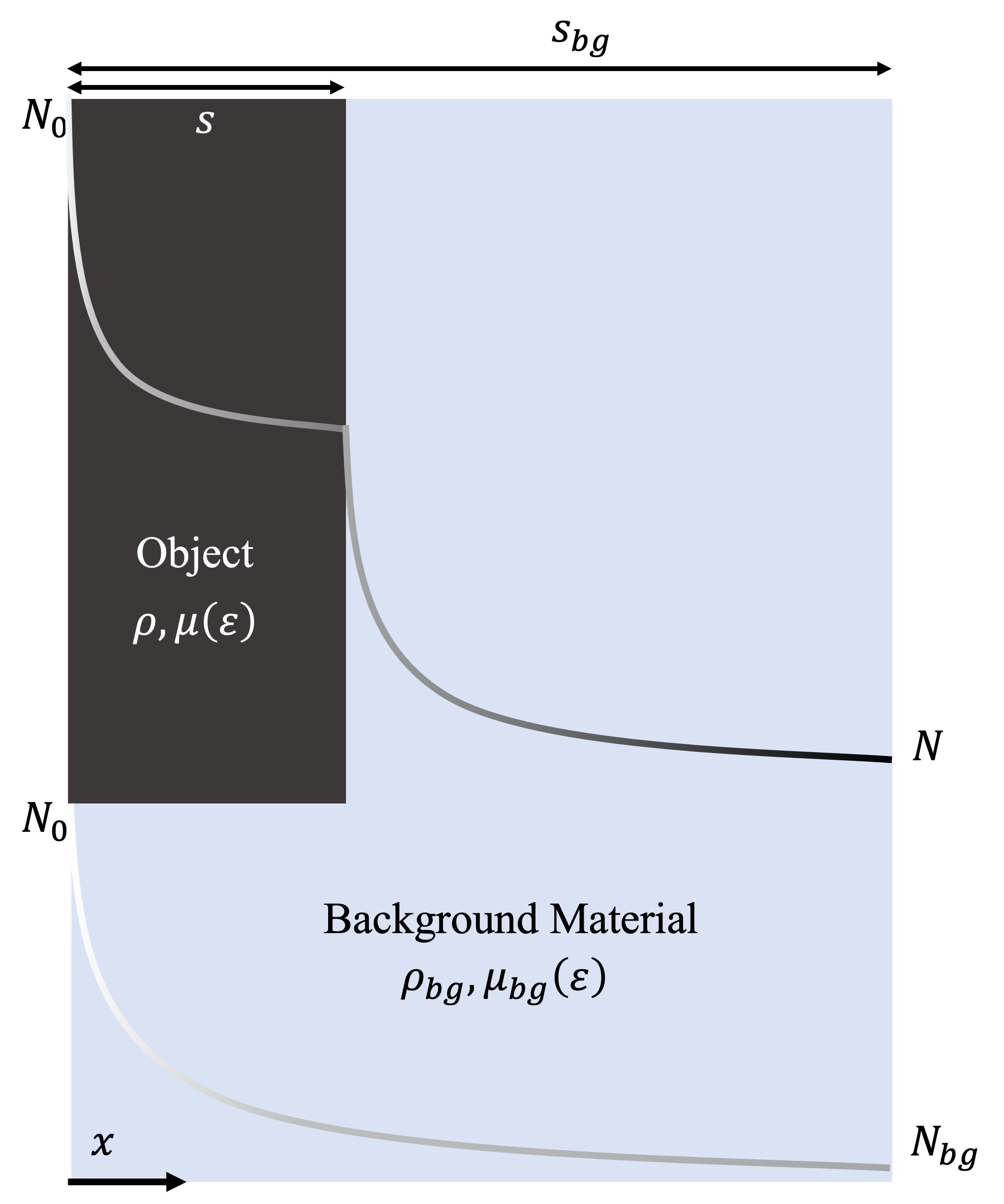}
    \caption{A simplified diagram depicting the Beer-Lambert law. As the photons pass through an object or the background material they are attenuated exponentially. While the object is shown as a solid slab for simplicity, naturally similar calculations are usable for a ray passing though part of a spherical flow tracer with equivalent thickness and composition.}
    \label{fig:BLdiag}
\end{figure}

Software such as in \cite{parker_experimentally_2022} can evaluate equation \ref{eq:BLlaw} for an arbitrary experiment design to produce a predicted image. More computationally expensive Monte Carlo simulations could include the effects of scatter and more for flow experiments where such effects rise to first order effects (e.g. liquid metal flows). For the purpose of elucidating general guidelines for designing in-lab XPV systems, though, we will proceed with `'back of the envelope" calculations and make some simplifying assumptions. Let us assume a two-material system. That is, only a single object with uniform thickness, density, and mass attenuation in a uniform background medium. Then, to calculate the number of photons detected at the object's location, equation \ref{eq:BLlaw} simplifies to
\begin{multline}\label{eq:BLobject}
    N = \int_{E_{thresh}}^{E_{max}} I_0'(\varepsilon) \eta(\varepsilon) A t_{exp} \exp \left[ - \rho \mu(\varepsilon) s - \rho_{bg} \mu_{bg}(\varepsilon) (s_{bg} - s) \right] d\varepsilon = \\
    \int_{E_{thresh}}^{E_{max}} I_0'(\varepsilon) \eta(\varepsilon) A t_{exp} \exp \left( -\rho_{bg} \mu_{bg}(\varepsilon) s_{bg} \right) \exp \left[ (\rho_{bg} \mu_{bg}(\varepsilon) - \rho \mu(\varepsilon) )s \right] d\varepsilon
\end{multline}
where $s$ is the distance through the object and $s_{bg}$ is the distance through the background. Similarly, the number of photons detected in the background is
\begin{equation}\label{eq:BLbg}
    N_{bg} = \int_{E_{thresh}}^{E_{max}} n_{bg}(\varepsilon) d\varepsilon = \int_{E_{thresh}}^{E_{max}} I_0'(\varepsilon) \eta(\varepsilon) A t_{exp} \exp \left( - \rho_{bg} \mu_{bg}(\varepsilon) s_{bg} \right) d\varepsilon,
\end{equation}
where $n_{bg}(\varepsilon)$ is the number of detected background photons with energy $\varepsilon$. Figure \ref{fig:BLdiag} depicts the simplified Beer-Lambert law diagrammatically.

We can now estimate the contrast. Combining equations \ref{eq:contrast}, \ref{eq:BLobject}, and \ref{eq:BLbg} results in
\begin{equation}\label{eq:abscont}
    C = \int_{E_{thresh}}^{E_{max}} n_{bg}(\varepsilon) \lvert 1 - \exp \left[ \left( \rho_{bg}\mu_{bg}(\varepsilon) - \rho \mu(\varepsilon) \right) s \right] \rvert d\varepsilon.
\end{equation}

\begin{figure}
    \centering
    \includegraphics[width=0.65\textwidth]{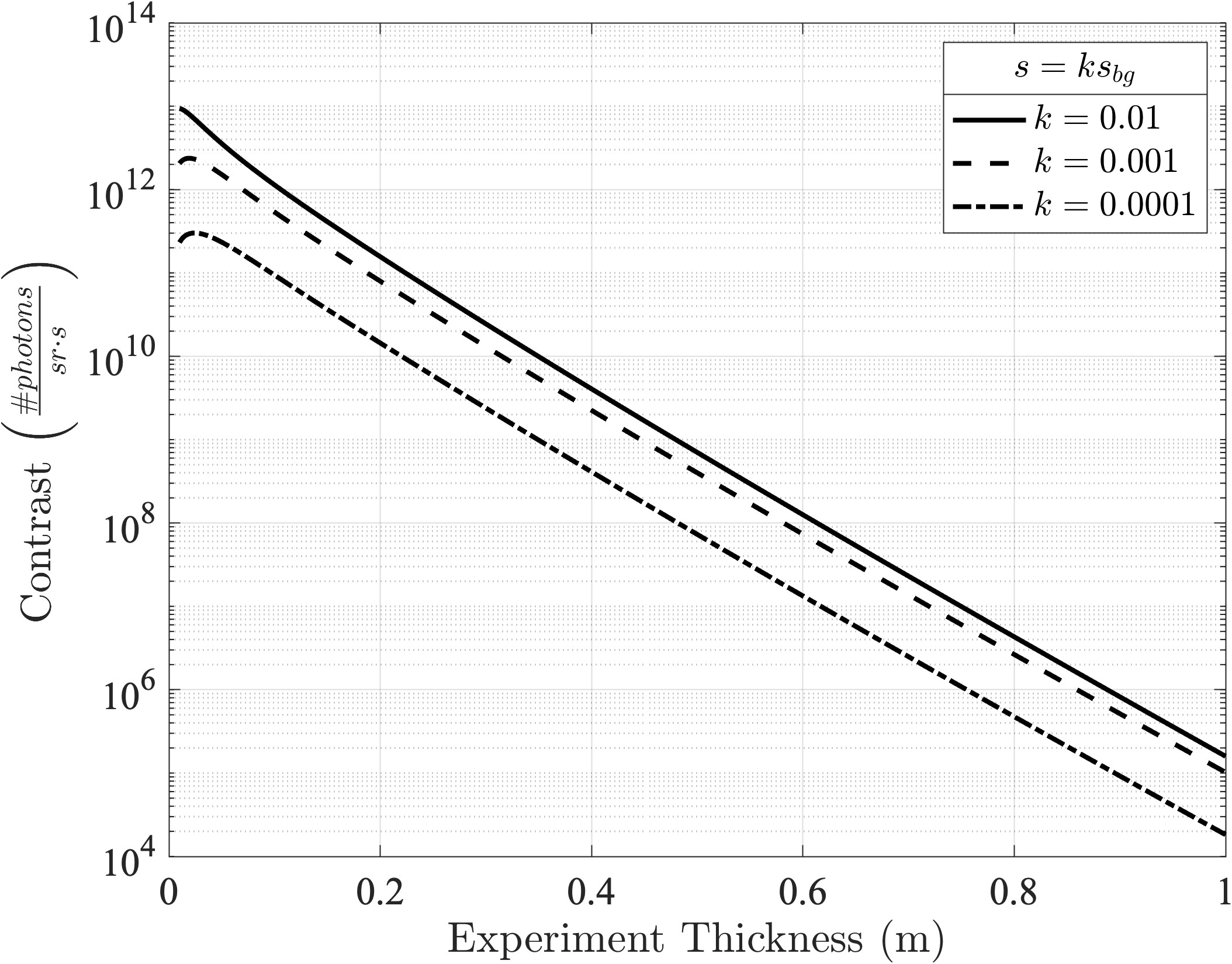}
    \caption{The contrast between silver (i.e., a hypothetical solid particle of silver) and water as a function of the total experiment thickness calculated with equation \ref{eq:abscont}. A MetalJet E1+ source spectrum at 160~kVp, 700~W and an XC-Thor are used. The detector energy threshold is 15~keV. A constant DQE of 0.75 is assumed. The depth of silver ($s$ in this plot) is a ratio, $k$, of the depth of water, ($s_{bg}$ in this plot).}
    \label{fig:expthickcont}
\end{figure}

Intuitively, materials with the largest differences in mass attenuation and density will generate largest contrast, as seen in the final exponential of equation \ref{eq:abscont}. Furthermore, a large object (i.e., $s$ is large) generates more contrast than a small one. Slightly less intuitive is that having a smaller domain (i.e. $s_{bg}$ is small) increases the image contrast. In other words, building the smallest feasible flow experiment maximizes the contrast by maximizing the number of detected photons, as shown in figure \ref{fig:expthickcont}. A smaller flow experiment has the additional benefit that one can increase the spatial resolution by increasing the geometric magnification while keeping the experiment in the FOV. Smaller flow experiments have additional benefits for reducing blurring artifacts, which will be described below. Nevertheless, the experiment size will need to be a balance between the desired flow regime and optimizing the X-ray image quality.

Lastly, choosing materials whose largest mass attenuation coefficient difference occurs at the photon energy dominant in the X-ray spectrum detected maximizes contrast, pursuant to other materials in the beam path that may attenuate some emitted photon energies prior to their arrival at the tracer and fluid. This is best demonstrated by example. Assume that we have a monochromatic source that either results 100 or 1,000 detected background photons, $N_{bg}$, at 10~keV or 30~keV, respectively. Without loss of generality let us also assume that all densities and material thicknesses are unity. At 10~keV let the mass attenuation difference be 0.2; at 30~keV the mass attenuation difference is 0.1. Although the mass attenuation difference is larger at 10~keV, according to equation \ref{eq:abscont} the absolute contrast will be larger at 30~keV because there are more photons at that energy.

\subsection{Signal to Noise Ratio}
Contrast-limited imaging refers to images in which the contrast does not vastly exceed the image noise. The relevant factor is naturally the SNR, defined here as
\begin{equation}\label{eq:snrdef}
    SNR \equiv \frac{C}{\sigma},
\end{equation}
where $\sigma$ is the standard deviation of the image background surrounding the object of interest. We can use the SNR to establish a lower bound on the exposure time, $t_{exp}$, required to capture useful images.

X-ray source photon emission, photon-medium interactions, and  photon arrival and detection processes are well modeled as a Poisson process. If we assume a Poisson noise distribution with the number of photons in a noiseless image being the expected value, then the standard deviation of the surrounding image noise is
\begin{equation}\label{eq:poisssigma}
    \sigma = \sqrt{N_{bg}}.
\end{equation}
However, assuming Poisson noise is overly idealistic in conditions where photon pile-up in PCDs begins to be non-negligible, and in systems with large amounts of scattered photons the standard deviation will be larger than in \ref{eq:poisssigma}. Typical scintillator-based or other energy-integrating detectors are often modeled with a combined Gaussian and Poisson distribution.

We can also define the relative contrast. The relative contrast tells us how strong the contrast is relative to the background and is defined as
\begin{equation} \label{eq:relcont}
    \Tilde{C} \equiv \frac{C}{N_{bg}}.
\end{equation}
Substituting equations \ref{eq:poisssigma} and \ref{eq:relcont} into equation \ref{eq:snrdef}, we obtain
\begin{equation}
    SNR = \Tilde{C}\sqrt{N_{bg}}.
\end{equation}
In other words, the SNR scales according to the square root of the number of detected photons. To establish the lower bound on the exposure time, we can decompose $\Tilde{C}$ and $N_{bg}$ into their rates and the exposure time, $\Tilde{C}' t_{exp}$ and $N_{bg}' t_{exp}$, respectively, to establish $SNR \geq 1$.
\begin{equation}\label{eq:exptime}
    t_{exp} \geq \left( \Tilde{C} \sqrt{N_{bg}'}' \right)^{-2/3}.
\end{equation}
Software such as Spekcalc \cite{poludniowski_calculation_2007, poludniowski_calculation_2007-1, poludniowski_spekcalc_2009} in combination with \cite{parker_experimentally_2022} can be used to estimate the necessary exposure time to achieve desired SNR. One can simulate a single image of arbitrary exposure time (or, equivalently, per unit time) and calculate the number of background photons and the relative contrast for that image. Using those two parameters, and the arbitrary exposure time that is simulated, one can evaluate equation \ref{eq:exptime} assuming the noise remains in the Poisson regime. This exposure time is the minimum required to be able to observe the particles and identify their location within an image. The upper exposure time limit is set by the maximum tolerable motion blur artifacts, which will now be discussed.

\subsection{Motion Blur Artifacts}\label{ssec:motblur}
Motion blur artifacts can occur in multiple forms depending on whether the imaging is 2D or 3D. There is projection motion blur, rotational motion blur, and tomographic motion blur. Projection motion blur refers to blurring within a single image exposure. Rotational motion blur refers to artifacts created while capturing TXPV data by rotating the source-detector pair relative to the flow experiment. Tomographic motion blur is caused by particle movement over the course of a single CT scan. How this motion blur manifests in the reconstructed volume depends on the reconstruction algorithm used. An in-depth discussion of this behavior is beyond the scope of the current study but will be discussed in subsequent publications.

\subsubsection{Projection Motion Blur}
Projection motion blur is straightforward to account for. In 2D-projected XPV, a particle will ideally not move farther than its diameter (or two) within a single exposure. With $U$ as the characteristic or maximum flow speed and $n$ as the number of permissible particle diameters of motion we can simply express the exposure time criteria as,
\begin{equation}
    t_{exp} \leq \frac{n d_p}{U}.
\end{equation}
We can use this exposure time maximum and the minimum set by the SNR requirement to determine the feasibility of the XPV experiment:
\begin{equation}
    \left( \Tilde{C} \sqrt{N_{bg}'}' \right)^{-2/3} \leq t_{exp} \leq \frac{n d_p}{U}.
\end{equation}

\subsubsection{Tomographic and Rotational Motion Blur}
When capturing TXPV data, one must consider particle motion over the course of the CT scan and due to rotation relative to the source-detector pair. If particles move too much during the CT scan, their motion will be warped and blurred. As with projection motion blur, we can set the limit to be
\begin{equation}\label{eq:tomoblur}
    U t_{CT} \leq n d_p,
\end{equation}
where $t_{CT}$ is the time it takes to complete one CT scan.

We can use equation \ref{eq:tomoblur} to establish a minimum speed at which we must rotate the source-detector pair and flow experiment relative to one another. This speed, $\omega$, is given by $2\pi/t_{CT}$. Thus,
\begin{equation} \label{eq:rotspeed}
    \omega \geq \frac{ 2\pi U }{ n d_p }.
\end{equation}
If the rotational motion is too quick, however, it can result in blurring within the projections. We refer to this as rotational blurring to distinguish from the fluid motion blurring discussed above. To avoid rotational blurring, we must set another limit
\begin{equation}\label{eq:rotblur}
    \omega R t_{exp} \leq n d_p,
\end{equation}
where $R$ is conservatively chosen to be the largest possible radial location of the tracer particles from the axis of rotation. Combined with the minimum rotation rate in inequality \ref{eq:rotspeed} required to avoid tomographic blur we can establish bounds on the necessary exposure time and rotation speed. Similar to projection blur above, our exposure time is bounded by the SNR requirement and the blurring requirements to be
\begin{gather}
    \left( \Tilde{C} \sqrt{N_{bg}'}' \right)^{-2/3} \leq t_{exp} \leq \frac{ n d_p }{ \omega R } \label{eq:exptimebound} \\
    \frac{2\pi U}{n d_p} \leq \omega \label{eq:rotspeedbound}
\end{gather}
Besides rotational blurring in the projections, if one is rotating the flow experiment instead of the source-detector pair then rotational accelerations are introduced that may influence the flow dynamics. The rotational acceleration may present an upper limit on inequality \ref{eq:rotspeed}.

\begin{figure}
    \centering
    \begin{subfigure}[b]{0.49\textwidth}
        \centering
        \includegraphics[width = \textwidth]{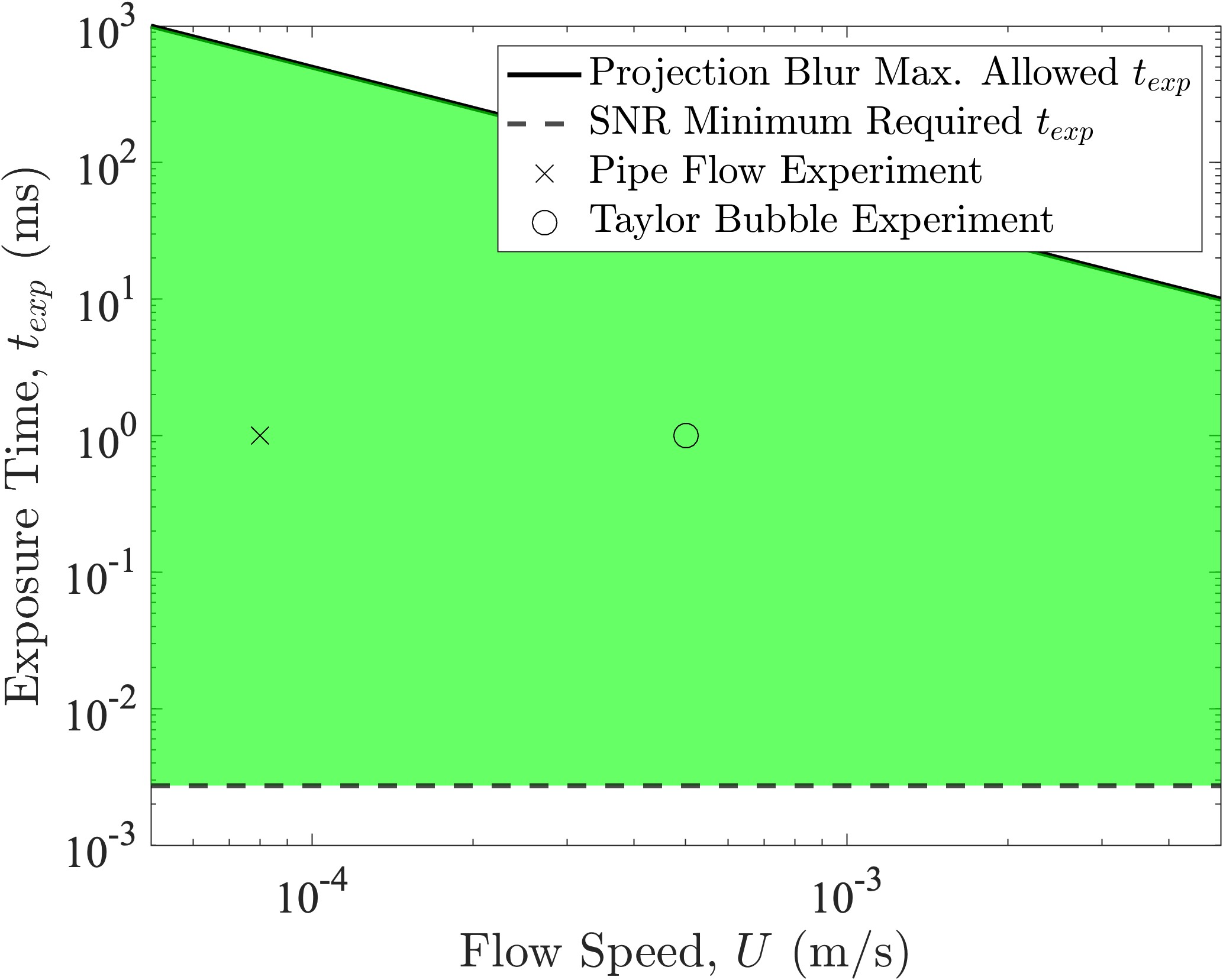}
        \caption{Projection blurring limits for one particle diameter of motion blur.}
    \end{subfigure}
    \hfill
    \begin{subfigure}[b]{0.49\textwidth}
        \centering
        \includegraphics[width = \textwidth]{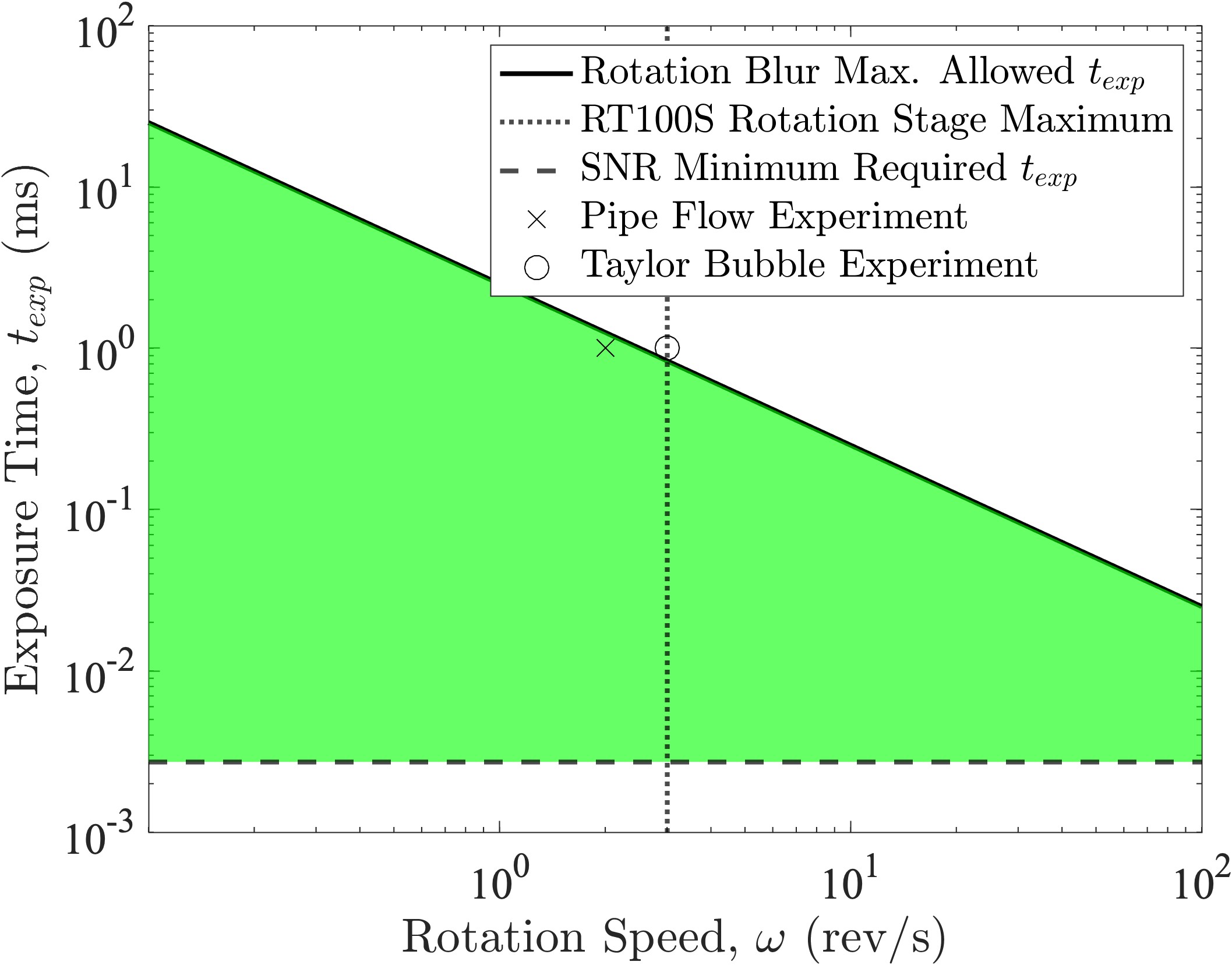}
        \caption{Rotation blurring limits for one particle diameter of motion blur.}
    \end{subfigure}
    \vfill
    \begin{subfigure}[b]{0.49\textwidth}
        \centering
        \includegraphics[width = \textwidth]{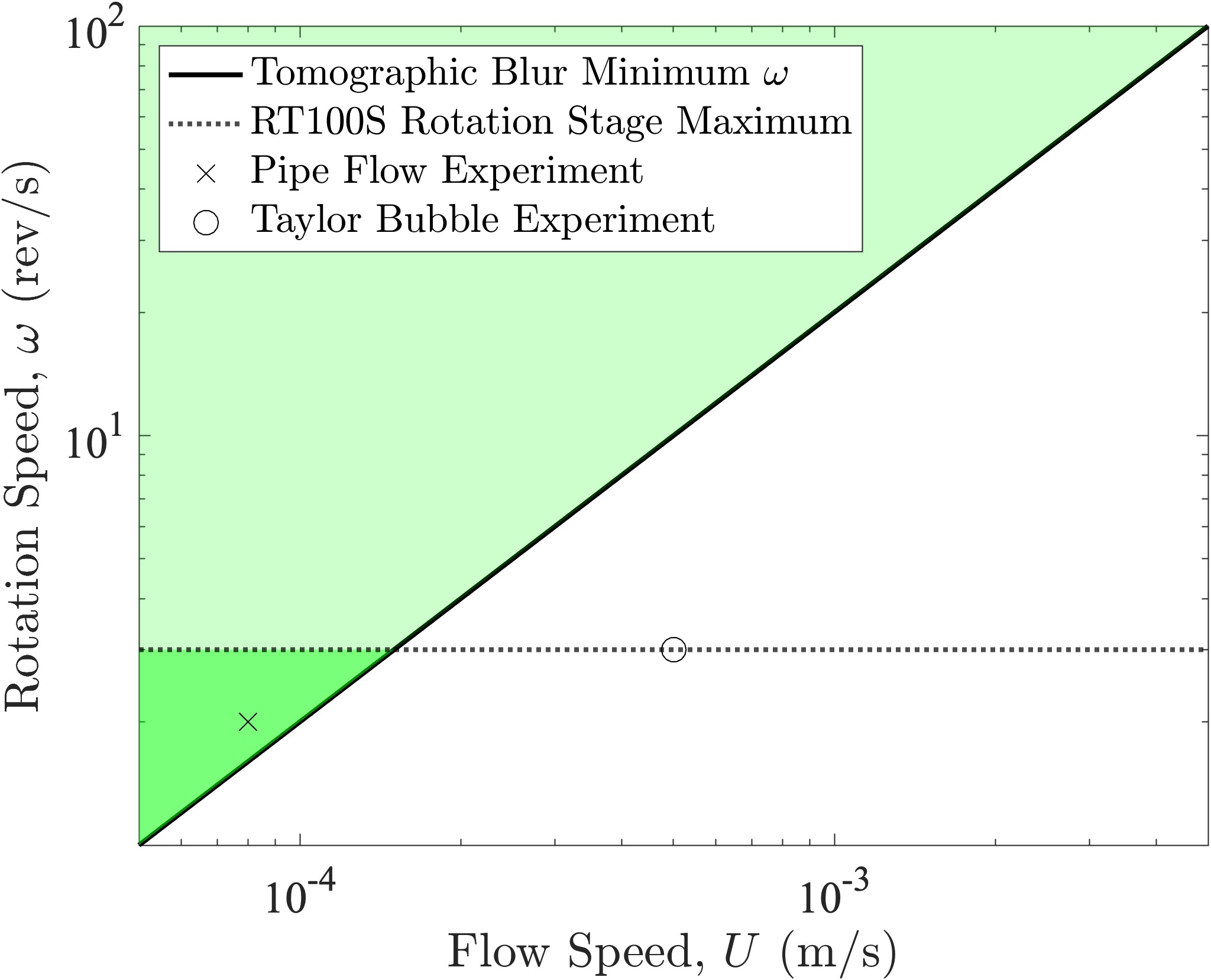}
        \caption{Minimum rotation speed for less than one particle diameter of tomographic blurring.}
    \end{subfigure}
    \caption{The exposure time and rotation speed bounds for a 50~\textmu m CW particle in a 6.35~mm diameter body of glycerine. Shaded regions indicate feasible range. The pipe flow and Taylor bubble experiment parameters are plotted. The particle contrast is calculated from the ray passing through $d_p$ and the ray passing through the background immediately adjacent to the particle. The detector and source settings are the same as figure \ref{fig:expthickcont}. Projection blurring presents a much less stringent requirement and greatly expands the range of flows that can be studied, motivating techniques to reduce the number of angles needed to localize the tracer particles.}
    \label{fig:time_speed}
\end{figure}

If it is not possible to satisfy inequality \ref{eq:exptimebound} and \ref{eq:rotspeedbound}, then the flow speed $U$, experiment size $R$, or both, must decrease. Alternatively, the particle diameter $d_p$ could be enlarged. This latter option is fraught however, as a larger tracer particle may not be an accurate flow tracer, as discussed in section \ref{ssec:tracers}. The tracer particles must have a diameter such that the Stokes number, $St \ll 1$. Larger particles are less accurate in large velocity gradients such as those encountered in boundary layers near walls. Often, regions with large velocity gradients are the regions of the most interest. Figure \ref{fig:time_speed} shows the limits in the parameter space where one achieves sufficient SNR but with at most one particle diameter of motion blurring based on inequalities \ref{eq:exptimebound} and \ref{eq:rotspeedbound}. As shown in figure \ref{fig:time_speed} the maximum flow speeds measurable with TXPV in even a small experiment such as this one are \textit{O}(1-10~mm/s) if one requires full 360 degree rotation. Reducing the number of angles needed for particle localization is the next step in improving the usefulness of in-lab TXPV such that one can approach the less stringent limits imposed by projection blurring.

The exposure time bounds presented here are ``fuzzy". Conservative assumptions are used throughout and the actual bounds may be more expansive or restrictive than the analytical bounds imply and with advanced post-processing more motion blur or less SNR may be tolerated in some situations. For example, flow at the largest radial extent of the experiment may not be particularly interesting (e.g., low velocity near a wall), so the requirement imposed in equation \ref{eq:rotblur} may be unnecessarily stringent. According to figure \ref{fig:time_speed} and inequality \ref{eq:rotspeed}, rotating at 580~dps is the minimum speed to resolve the flow and avoid motion blur for the TXPV experiments shown here. Nevertheless, we rotate the experiment at 720~dps for a higher temporal resolution and do not observe notable motion blurring to be occurring. On the other hand, the lower bound on exposure time assumes an idealized noise intensity for a PCD. Realistically this lower bound may be higher, particularly if there is substantial photon pile-up.

The procedure described so far to determine the feasibility of TXPV experiments for a given flow experiment does not consider spatial resolution and instead focuses on temporal resolution limitations. Spatial resolution limitations of course important to consider, however. The spatial resolution must be high enough, for example, that one can resolve the tracer particles. At present, though, temporal limitations are typically the stronger limiting factor for in-lab TXPV use for flows of interest.

\section{Conclusions} \label{sec:conc}
Measurements of Poiseuille pipe flow, a rising Taylor bubble, and a laminar jet demonstrate the improved capabilities of in-lab high speed 2D XPV, TXPV, and CT achievable with a LMJ source, X-ray imaging-specific tracer particles, and a PCD. The 2D and 3D Poiseuille pipe flow measurements demonstrate the improved accuracy of the technique compared to our prior work. 2D-projected Poiseuille pipe flow data is acquired at 1~kHz, showing that \textit{O}(1~kHz) acquisition rates are possible with \textit{O}(1-10~cm) domains in the laboratory with the requisite hardware. The Taylor bubble and KI jet measurements demonstrate the applicability of TXPV to multiphase flows and the ability of this technique to reconstruct the flow morphology in 3D. The laminar KI jet experiments demonstrate the potential for high speed CT to capture scalar mixing in 3D alongside 3D flow velocity measurements.

The measurements captured in this study are an order of magnitude faster than the previous state of the art thanks to the combined usage of a PCD, X-ray imaging-specific \textit{O}(50~\textmu m) tracer particles, and a LMJ X-ray source. This combination enabled imaging frame rates of 1~kHz, which is 15$\times$ faster than previous comparable studies. Additionally, the bright LMJ source allows us to capture full 360 degree data for TXPV reconstructions at nearly an order of magnitude higher speed than previous studies. These acquisition speeds bring more diverse fluid dynamic applications -- from biological flows, to soil mechanics, to boiling flows -- into the realm where in-lab XPV may be used.

Many of the limitations in this study are practical in nature, and are easily overcome. The rotation stage used here, for example, is limited to 1080~dps, meaning 3~Hz CT was the fastest achievable scan speed due to the rotation stage. Additionally, there is only a single, static source-detector pair. More source-detector pairs would improve the temporal resolution by requiring fewer angles to be traversed to capture a full CT scan. The detector used in this study also has a relatively small buffer size, which limited the number of frames that could be continuously acquired. This is to say that there is ample room for improvement that we can expect to be readily achievable.

A more involved improvement to the X-ray imaging setup could include implementing SXPV, which uses multiple static source-detector pairs. For example, three source-imager pairs would enable particle or feature tracking at imaging frame rates, i.e., \textit{O}(1~kHz). Such a system would theoretically only be temporally limited by the detector frame rate and source brightness. In this case, as demonstrated by the 2D-projected pipe flow measurements, that would allow 1~kHz SXPV. Source and detector technology is rapidly improving, so future source-detector pairs will likely be able to go to even faster speeds. TXPV is also possible with a single source-detector pair by using limited-angle reconstruction, although that is not attempted in this study and is an area for future work.

A procedure for designing a high speed TXPV experiment is presented based on limitations arising from contrast as estimated with the Beer-Lambert equation. This procedure will hopefully encourage greater adoption of TXPV and XPV experiments in general within the fluid dynamics community. 

XPV is rapidly improving. Already these techniques could be useful for slow flows such as porous media flows and biological flows. The authors expect that the technique will see a dramatic increase in data acquisition speed over the coming years as PCDs become faster, sources become brighter, and newer, better X-ray tracer particles are developed. This study shows that the combination of these three elements yields orders of magnitude improvement in acquisition speed, thereby increasing the usefulness of XPV.

\section*{Acknowledgments}
We gratefully acknowledge the support of NSF EAGER award \#1922877 program managers Ron Joslin and Shahab Shojaei-Zadeh and the additional support provided by the Society of Hellman Fellows Fund. This work was also partially supported by AFRI Competitive Grant no. 2020-67021-32855/project accession no. 1024262 from the USDA National Institute of Food and Agriculture. This grant is being administered through AIFS: the AI Institute for Next Generation Food Systems (\url{https://aifs.ucdavis.edu}).

\appendix

\section{Image Processing Algorithms}\label{app:A}

\subsection{2D Pipe Flow}\label{app:Apipe}
For the 2D pipe flow experiments, the following algorithm was applied in LaVision DaVis version 8.4:
\begin{enumerate}
    \item Apply a mask such that only the pipe inner diameter is visible.
    \item Subtract the average intensity of all the frames from each frame.
    \item Apply a strict sliding minimum filter with a scale length of 5 pixels.
    \item Subtract a constant 0.01 from each frame.
    \item Set all pixel values below zero to zero.
    \item Apply a median filter with a $5 \times 5$ kernel size.
    \item Multiply each frame by a constant 100.
    \item Conduct PIV in two iterations:
        \begin{enumerate}
            \item One pass of $64 \times 64$ pixel window size with 50\% overlap.
            \item Two passes of $48 \times 48$ pixel window size with 50\% overlap.
            \item Apply vector removal based on pixel displacement with the following limits: $V_x \in [-5, 5]$, $V_y \in [-12, 0]$.
        \end{enumerate}
    \item Conduct PTV with particle tracking assisted with PIV. The particle size range is 1 -- 8 pixels; the intensity threshold is 0 counts; the correlation window size is 32 pixels; the allowed vector range relative to reference is 20 pixels.
    \item Export to MATLAB to apply calibration and plot results. Average the profile over 850 frames since PTV requires statistical convergence to calculate a profile.
\end{enumerate}

\subsection{3D Pipe Flow and Taylor Bubble Flows}\label{app:A3Dexper}
In the tomographic, 3D flow experiments, the following was algorithm was used. First, for the CT reconstruction:
\begin{enumerate}
    \item Convert flat field corrected images to integrated attenuation values by dividing by the empty field of view (FOV) pixel value.
    \item Identify the center of rotation.
    \item Use normalization ring artifact removal with a standard deviation of 15 pixels.
    \item Use ASTRA Toolbox (\cite{van_aarle_astra_2015, aarle_fast_2016}) to reconstruct the volume. A standard Ram-Lak filter is used.
\end{enumerate}

For the segmentation, MATLAB is used to segment the reconstructed data based on the voxel intensity in the following manner. For each reconstructed volume:
\begin{enumerate}
    \item Normalize the reconstructed volume to grayscale.
    \item 3D median filter the volume with a $5 \times 5 \times 5$ kernel and replicating boundary conditions.
    \item Binarize the volume using \texttt{imbinarize}. The threshold is adaptive, the sensitivity is 55\%, and the foreground polarity is dark.
    \item Invert the binarized image.
    \item Use \texttt{conncomp} to identify 6-connected components volumes.
    \item Use regionprops3 to measure the volume and centroid location of the volumes. Weight the centroid location with voxel intensity from the normalized reconstructed volume.
\end{enumerate}

For particle tracking, a MATLAB code was developed based on a nearest-neighbor algorithm. The code will be publicly available at the FLOW Lab website: \texttt{flow.berkeley.edu}.

\subsection{KI Jet Processing}\label{app:KIjet}
The KI jet is captured with high speed CT. These scans were reconstructed using ASTRA Toolbox (\cite{van_aarle_astra_2015}). Each reconstruction consisted of 250 projections. For better temporal interrogation, we reconstruct each volume with a 90\% overlap in the projections used. For example, the first reconstruction is done with projections 1 to 250; the second reconstruction is done with projections 25 to 275, and so on.

Post-reconstruction, MATLAB is used to correct the source flicker and calculate the solution concentration. The algorithm is as follows:
\begin{enumerate}
    \item The average intensity of the middle slice of each volume is normalized by the average intensity of the middle slices for all volumes. This provides the flicker correction factor.
    \item Each volume is divided by the correction factor, then 3D Gaussian-filtered with an $11 \times 11 \times 11$ kernel with a standard deviation of 7 pixels.
    \item The first five and the final five volumes are averaged to calculate the zero concentration and fully mix concentration voxel intensities. The average inner pipe voxel intensity is taken to be the intensity for its respective concentration. The coefficients in equation \ref{eq:concen} are then calculated.
    \item Equation \ref{eq:concen} is applied to each Gaussian-filtered volume to calculate the concentration in each voxel.
\end{enumerate}

\printbibliography

\end{document}